\documentclass[%
reprint,
amsmath,amssymb,
aps,
pra,
]{revtex4-1}

\usepackage{graphicx}
\usepackage{dcolumn}
\usepackage{bm}

\usepackage{braket}
\usepackage{xcolor}
\usepackage{csquotes}
\usepackage{verbatim}
\usepackage{bbold}

\newcommand{\tr}{\mathrm{tr}}
\usepackage{hyperref}

\usepackage{float}

\begin{document}

\preprint{APS/123-QED}

\title{Local ergotropy and its fluctuations across a dissipative quantum phase transition}

\author{G. Di Bello$^{1,2,*}$}\author{D. Farina$^{1,2,*}$}\author{D. Jansen$^{2}$}\author{C. A. Perroni$^{3,4}$} \author{V. Cataudella$^{3,4}$}\author{G. De Filippis$^{3,4}$}
\affiliation{$^{1}$Dip. di Fisica E. Pancini - Università di Napoli Federico II - I-80126 Napoli, Italy}
\affiliation{$^{2}$ICFO-Institut de Ciencies Fotoniques, The Barcelona Institute of Science and Technology, 08860 Castelldefels (Barcelona), Spain}
\affiliation{$^{3}$SPIN-CNR and Dip. di Fisica E. Pancini - Università di Napoli Federico II - I-80126 Napoli, Italy}
\affiliation{$^{4}$INFN, Sezione di Napoli - Complesso Universitario di Monte S. Angelo - I-80126 Napoli, Italy}
\affiliation{$^*$Corresponding authors: G. Di Bello, grazia.dibello@unina.it and D. Farina, donato.farina@unina.it}



\begin{abstract}
We investigate a two-qubit open Rabi model, focusing on local ergotropy—the maximum extractable work by acting solely on the two qubits—within a parameter regime where a Berezinskii-Kosterlitz-Thouless dissipative phase transition occurs. First, we aim to define a protocol for charging, storing, and discharging the two-qubit system, interpreted as the working principle of an open quantum battery. Second, we examine the impact of the phase transition on ergotropy and identify potential markers. To achieve these goals, we construct an ad-hoc charging unitary operator, leveraging our knowledge of the ground state near the transition to bring it into a decoherence-free state during storage. Using state-of-the-art numerics based on matrix product state representation, we reveal that high couplings to an external bath approximately double the local ergotropy immediately post-charging. Over time we observe oscillatory behaviors in ergotropy and its fluctuations, which undergo significant changes near the transition, signaling its occurrence. Furthermore, we optimize local ergotropy over time using a physically inspired ansatz, enabling work extraction at a generic time (local ergotropy never reaches zero). Our work proposes a tunable, experimentally realizable protocol for work extraction, leveraging decoherence-free states and phase transitions. Additionally, it sheds light on the complex interaction between local ergotropy and quantum phase transitions.

\end{abstract}

\maketitle

\section{Introduction}
Ergotropy quantifies the maximum extractable average work from a closed quantum system under cyclic protocols \cite{allahverdyan2004maximal}. For a system of dimension $d$, in a state $\rho$ and described by a Hamiltonian $H$, ergotropy can be defined as 
\begin{equation}
    \mathcal{E}=\max_{U \in \mathcal{U}(d)} \left\{\tr(H \rho)-\tr(U \rho U^\dag H)\right\}\,,
    \label{ergo}
\end{equation}
where $\max$ denotes the maximum taken over all unitary transformations $U$ in the set of $d$-dimensional unitary operators $\mathcal{U}(d)$, which act on the entire system to extract work from it, and $\tr$ is the trace over the whole system. 

When the system $S$ is in contact with another system $E$ (environment) on which we do not have unitary control, one can define different ergotropic quantities. Hence, let the $SE$ Hamiltonian be of the (general) form 
\begin{equation}
    H_{SE}=H_{S}+H_{E}+V_{SE}\,,
    \label{total-ham}
\end{equation}
where $H_S$ and $H_E$ are local terms on $S$ and $E$ and $V_{SE}$ is the interaction term. It is important to note that $S$ can be viewed as an open quantum battery, as it interacts with the environment (open) and serves as the working principle from which one aims to extract work (quantum battery). 

Given this context, there is increasing interest in studying the effects of the environment on work extraction. This includes both Markovian and non-Markovian effects, as discussed in \cite{choquehuanca2024dynamics}, and the environment's influence on the remote charging of a quantum battery \cite{song2024remote}. Additionally, experimental works have presented possible implementations of many-body quantum batteries in superconducting devices \cite{yang2023resonator}.

To deal with the presence of the environment, different strategies have been proposed. On one hand, we can think of first \textit{instantaneously} turning off the coupling between $S$ and $E$ through a Hamiltonian quench (switch-off, \enquote{$so$}, protocol), paying an energy price
\cite{barra_thermodynamic_2015,ito_fundamental_2017,andolina_charger-mediated_2018,chiara_reconciliation_2018,strasberg_quantum_2017}
\begin{equation}
\Delta_{so}=-\tr(V_{SE} \rho_{SE})  
\label{so-delta}
\end{equation}
and then from the isolated system $S$ extract the maximum work, using the unitary operator corresponding to the ergotropy of the \textit{subsystem} $S$,
\begin{equation}    \mathcal{E}_{sub}=\mathcal{E}(\rho_S, H_S)\,,
\label{sub-ergo}
\end{equation}
where $\rho_S=\tr_E(\rho_{SE})$ is the reduced density matrix of $S$.

The maximum extractable work through this procedure is then called the switch-off ergotropy
\begin{equation}
    \mathcal{E}_{so}
    =\mathcal{E}_{sub}-\Delta_{so}\,.
    \label{so-ergo}
\end{equation}
On the other hand, in presence of the coupling $V_{SE}$, we can directly extract work via local unitary transformations on $S$. The concept of local ergotropy \cite{LErgo} formalizes this procedure. 
It is defined as 
\begin{equation}
\mathcal{E}_{S}=
\max_{U_S \in \mathcal{U}(d_S)} \left\{\tr(H_{SE} \rho_{SE})-\tr(U_S \rho_{SE} U_S^\dag H_{SE})\right\}\,,
    \label{local-ergo}
\end{equation}
where now the maximization is restricted to unitary operations that are local on $S$.
This automatically implies that local ergotropy is upper-bounded by the global ergotropy of the compound $SE$, i.e.
$\mathcal{E}_{S}\leq \mathcal{E}(\rho_{SE}, H_{SE})$.
Notably, in several cases local ergotropy implies an advantage in work extraction when compared with the switch-off protocol \cite{LErgo} that can be further improved when properly exploiting the \textit{free} dynamics of the $SE$ compound \cite{castellano2024extended}. Here, we compare local ergotropy with the switch-off protocol as in \cite{LErgo}, but we consider a realistic open system with a bath and perform work extraction at a generic time, since local ergotropy never reaches zero. 

It is worth noting that, since we aim to exploit the environment dynamics to extract work from the system, we need to study the time-dependent behavior of the $SE$ compound. One approach to this problem is to use Lindblad master equations to describe the dynamics of the open quantum battery \cite{PhysRevB.99.035421}. However, their validity is well known to be restricted by Markovian and weak-coupling assumptions \cite{breuer2002theory}. To obtain a rigorous treatment, it is necessary to consider the full unitary dynamics of the $SE$ system. This is achieved through state-of-the-art numerical simulations.

In the recent years connections between ergotropic quantities and many-body phenomena are attracting increasing interest \cite{rossini2019many,rossini2020quantum,zhao2021quantum, zhao2022charging, 
barra2022quantum,arjmandi2023localization,grazi2024enhancing, feliu2024system,gyhm2024beneficial}.
Notably, it has been discovered that driving a system across a quantum phase transition (QPT) can enhance the energy stored in an XY spin chain quantum battery \cite{grazi2024enhancing}. Moreover, in a nearest-neighbor spin chain, ergotropy and energy storage exhibit a critical point reminiscent of the one dictated by the QPT, beyond which they significantly increase.

In this framework, it is particularly relevant to understand if QPT can be beneficial for extracting work from open systems and, conversely, if ergotropic quantities can be used as witness of many-body phenomena in open systems. In particular, an analysis in this regard concerning the recently introduced notion of local ergotropy is still missing in the literature, even though some studies about work distribution across a QPT already exist \cite{RevModPhys.96.031001,niu2024work}. Filling this gap in literature, represents the first objective of our investigation. We will focus on an environment-induced QPT belonging to the class of strong coupling Berezinskii-Kosterlitz-Thouless (BKT) transitions.

Another aspect we will analyze is the effect of the decoherence-free subspace (DFS) in engineering our open quantum battery. A DFS is an eigenstate of our system $S$ that remains protected from environmental effects during its dynamics. We will attempt to protect the system from the environment by exploiting the DFS, while also leveraging the interaction with the environment to extract more work from the system. Similar attempts can be found in the literature, such as the work by Liu et al. \cite{liu2019loss}, which proposed the realization of a loss-free quantum battery by considering subspaces that are invariant with respect to the dynamics induced by the total Hamiltonian. However, extending this approach to many-body systems is not trivial. In this work, we provide such an extension in a specific model. Actually, as pointed out in \cite{RevModPhys.96.031001}, developing a systematic approach to identify robust and energetically favorable relaxation-free subspaces in scalable many-body systems is crucial.

Specifically, we numerically study the behavior of local ergotropy and its fluctuations across the BKT QPT occurring in a two-qubit open quantum Rabi model. It has also been demonstrated that within the same model, by designating one qubit as the battery and the other as the charger, the presence of a QPT can be leveraged to achieve enhanced working performance \cite{crescente2024boosting}. 

The second aim of this work is to propose a realistic protocol, describing charging, quasi-decoherence free storage and work extraction. We observe that the coupling to the bath is not detrimental to the local ergotropy; on the contrary, it increases as the coupling increases. By employing a local unitary gate based on the form of the ground state after the phase transition, we can approximately double the extractable work.

We also demonstrate that the dynamics of local ergotropy and its fluctuations changes behavior once the QPT has occurred in the model. The features of this behavior, primarily due to the bath evolution, can be controlled by adjusting the local unitary gate used for work extraction.

The paper is organized as follows.
In Sec.\,\ref{sec:model} we introduce the 
full Hamiltonian description of the system-environment compound, emphasizing relevant many-body phenomena characterizing the model, and presenting a concrete proposal for charging, storage and work extraction protocols. Section \ref{sec:results} presents our original numerical results regarding the relationship between ergotropic quantities and many-body phenomena. We describe the many-body state using a matrix product state (MPS) representation and employ density matrix renormalization group (DMRG) \cite{white_92,schollwoeck_2011} simulations to determine the ground state of the Hamiltonian. Subsequently, we evolve this state using time-dependent variational principle (TDVP) \cite{haegeman_11,haegeman_16} numerical simulations. We finally draw our conclusions in Sec.\,\ref{sec:conc} together with possible future developments.

\section{The model}
\label{sec:model}
\subsection{Hamiltionian description}
We consider a two-qubit quantum Rabi model, that is two interacting qubits connected through a harmonic oscillator to an Ohmic bath \cite{di2024optimal,di2023environment}. We set $\hbar=1$.
The terms characterizing the Hamiltonian \eqref{total-ham} of the $SE$ compound read as follows. The two-qubit system Hamiltonian $H_{S}$ is defined as:
\begin{equation}\label{eq:hsyst}
H_{S} = -\frac{\Delta}{2}(\sigma_x^1 + \sigma_x^2) + \frac{J}{4}\sigma_z^1\sigma_z^2\,,
\end{equation}
where $\Delta$ is the frequency of the two qubits, $J$ is the strength of the interaction between them, and $\sigma_i^j$ (with $i=x,y,z$ and $j=1,2$) are the Pauli matrices. The environment Hamiltonian and its interaction with the system are given by:
\begin{align}\label{eq:hE}
H_{E} &= \sum_{i=1}^{N} \omega_i b^{\dagger}_i b_i \\ \label{eq:VSE}
V_{SE} &= (\sigma_z^1+\sigma_z^2)\sum_{i=1}^N \lambda_i(b_i+b^{\dagger}_i)\,.
\end{align}
The bath is represented as a collection of $N$ oscillators with frequencies $\omega_i$, and creation (annihilation) operators $b^{\dagger}_i$ ($b_i$), interacting with coefficients $\lambda_i$ through a two-qubit magnetization-oscillator position coupling $g$. This form of interaction with the bath represents the Caldeira-Leggett model \cite{CaldeiraLeggett81,weiss2012quantum} for dissipation and can be experimentally realized by coupling two flux qubits to a resonator, which is further coupled to an Ohmic bath \cite{magazzu2019transmission,goorden2004entanglement}. This results in the definition of the bath spectral density $J(\omega)=\sum_{i=1}^{N}\left|\lambda_i\right|^2\delta(\omega-\omega_i)$ in the continuum limit:
\begin{equation}
\label{eq:J}
J(\omega)\xrightarrow[N\rightarrow\infty]{}\frac{2g^2\omega_0^2\alpha\omega}{(\omega^2-\omega_0^2-h(\omega))^2+(\pi\alpha\omega_0\omega)^2}\Theta\left(\frac{\omega_c}{\omega}-1\right).
\end{equation}
Here $\omega_0$ is the oscillator frequency that appears only in the form of the coupling coefficients $\lambda_i$, $\alpha$ is the strength of its interaction with the bath, $g$ is the coupling with the qubits, $h(\omega)=\alpha\omega_0\omega\log\left[\frac{\omega_c+\omega}{\omega_c-\omega}\right]$ and $\Theta(x)$ the Heaviside function describing the cutoff  (see \cite{de2023signatures} for a detailed derivation of this spectral density in the case of a single qubit interacting with an oscillator, or \cite{di2023environment} for the two qubits case). We emphasize that the spectral density is Ohmic at low frequencies: $J(\omega) \approx \frac{2 g^2 \alpha}{\omega_0^2}\omega$. Therefore, each qubit is coupled to the same oscillator bath through an effective constant proportional to $g^2 \alpha/ \omega_0^2$. This low-frequency behavior suggests the presence of a BKT QPT. Indeed, it has been proved \cite{de2023signatures} that, after the elimination of the bosonic degrees of freedom, the system is mapped to a model of two classical spin chains ferromagnetically interacting with each other. This coupling, at long distances, exhibits a power-law behavior of $1/r^2$, determined by the linear low-frequency regime of $J(\omega)$.

We emphasize that this model depicts a realistic scenario wherein a bath interacts with two qubits through an oscillator featuring a Rabi interaction. A related model has been proposed in Ref. \cite{LErgo}, where a single qubit interacts with a single cavity through a Jaynes-Cummings coupling. This aims to describe a model in which local ergotropy is a better choice than the global one and the switch-off one, with this superiority increasing as the coupling with the oscillator becomes stronger. The time evolution of that model shows periodic oscillations of local ergotropy, with minima occurring at zero. In our model, the presence of $N$ harmonic oscillators evolving with their phases over time deeply alters the amplitude of the local ergotropy and its fluctuations oscillations. Furthermore, the minima and maxima depend on the interaction strength $J$. As we will see later, choosing a high value for $J$, as in this work, ensures that we can always extract work, never reaching a state of zero local ergotropy.

\paragraph*{Decoherence-free state.---}
In what follows it will be useful to look
at the eigenstates of the Hamiltonian $H_S$ of the system alone. By diagonalizing the closed Hamiltonian $H_S$, the resulting spectrum reads $\frac{E\left(\frac{J}{\Delta}\right)}{\Delta}=\left\{-\frac{1}{4}\sqrt{\left(\frac{J}{\Delta}\right)^2+16};\,\frac{J}{4\Delta};\, -\frac{J}{4\Delta};\,\frac{1}{4}\sqrt{\left(\frac{J}{\Delta}\right)^2+16}\right\}$ and the corresponding eigenstates can be written in the Bell basis $\left\{\ket{S}\equiv\ket{\Psi^-},\,\,\ket{TAFM}\equiv\ket{\Psi^+},\,\,\ket{TFM+}\equiv\ket{\Phi^+},\right.$
$\left.\ket{TFM-}\equiv\ket{\Phi^-}\right\}$, 
where $S$ stands for ``singlet", while $T$ for ``triplet", and the specifications $AFM$ and $FM$, respectively, for ``antiferromagnetic" and ``ferromagnetic".
Hence, the Hamiltonian eigenstates read:
\begin{align}
\label{eigen-closed}
    \ket{0}=&\,a\left(\frac{J}{\Delta}\right)\ket{TAFM}-b\left(\frac{J}{\Delta}\right)\ket{TFM+}    
    \\
    \nonumber
    \ket{1}=&\ket{TFM-}\\
    \nonumber
    \ket{2}=&\ket{S}\\
    \nonumber
    \ket{3}=&\,a\left(\frac{J}{\Delta}\right)\ket{TFM+}+b\left(\frac{J}{\Delta}\right)\ket{TAFM},
\end{align}
where $0,1,2,3$ go from the ground state ($0$) to the most excited one ($3$). The two coefficients $a$ and $b$ depend only on the parameters of the Hamiltonian ($J/\Delta$) and are defined as follows:
\begin{align}
\label{eq:afac}
    a\left(\frac{J}{\Delta}\right)&=\frac{4}{\sqrt{16+\left(-\frac{J}{\Delta}+\sqrt{16+\left(\frac{J}{\Delta}\right)^2}\right)^2}}\\
    \label{eq:bfac}
    b\left(\frac{J}{\Delta}\right)&=-\frac{\sqrt{16+\left(\frac{J}{\Delta}\right)^2}-\frac{J}{\Delta}}{\sqrt{16+\left(-\frac{J}{\Delta}+\sqrt{16+\left(\frac{J}{\Delta}\right)^2}\right)^2}}.
\end{align}
It is worth noting that the singlet state $\ket{S}$ serves as the DFS in our model. This is because $\ket{S}_S$ is an eigenstate of the closed Hamiltonian $H_S$ and
$\ket{S}_S\otimes \ket{\vec{n}}_E$ ($\ket{\vec{n}}_E$ being a generic Fock state of the multiboson system) is an eigenstate of the entire Hamiltonian $H$. 
In particular, the evolution induced by the interaction Hamiltonian $V_{SE}$ does not affect the singlet state,
as 
\begin{equation}
    V_{SE} \ket{S}_S\otimes\ket{B}_E=0\,,
\end{equation}
for any state $\ket{B}_E$ of the bath.

\subsubsection{Phase diagram of the model}
The model we have chosen to implement our work extraction protocol has been investigated in previous works, revealing some interesting features. In particular, some of us showed \cite{di2023environment,de2023signatures} that the same model with two qubits (one in \cite{de2023signatures}) undergoes a BKT phase transition at zero temperature. This is a ground-state QPT induced by the presence of dissipation. We found that the BKT transition occurs with increasing qubits-oscillator coupling $g$, while fixing the ferromagnetic interaction between the qubits at $J=-10\Delta$. At a critical $g_c\approx 0.5 \Delta$, the favored interaction becomes ferromagnetic and the ground state is degenerate. 

Moreover, since we are also interested in the time behavior and want to find the unitary operator that makes the local ergotropy less dependent on time, or at least decreases it less over time, it's worth recalling that this model also undergoes a dynamical quantum phase transition (DQPT), as analyzed in \cite{di2023environment}. This means that a particular critical time $t_c$ can be found at which the Loschmidt's echo rate shows a kink, solely as a result of the coupling to the bath. The Loschmidt's echo quantifies the probability of the system returning to its initial state after a quench in the parameters of the Hamiltonian, followed by its evolution under the quenched Hamiltonian. This happens for coupling $g$ values in the same range as the thermodynamic QPT \cite{di2023environment}, although there is generally no exact understanding of how they are related \cite{heyl2018dynamical}.

We exploit the presence of the QPT in this system to design a local unitary gate for charging the two-qubit system. Our approach involves examining the ground state and choosing a strategy for local manipulation to induce charging. When the coupling to the bath is zero, the ground state is precisely $\ket{0}\otimes\ket{0,\dots,0}$, where $\ket{0}$ is the ground state of the closed two-qubit system in \eqref{eigen-closed}. However, after the BKT QPT occurs, the system must be in one of the two degenerate states, namely $\ket{\uparrow\uparrow}$ or $\ket{\downarrow\downarrow}$, along with a configuration of the bath modes \cite{de2023signatures,di2023environment}. On the other hand, near the QPT, it can be expressed as a linear combination of these two possibilities:
\begin{equation}
\label{gndstate}
\ket{\psi_{GS}}=
\alpha\ket{\uparrow\uparrow}\otimes\ket{B_{\uparrow}}+\beta\ket{\downarrow\downarrow}\otimes\ket{B_{\downarrow}}\,,
\end{equation}
where $\ket{B_{\uparrow}}$ and $\ket{B_{\downarrow}}$ represent potentially complex combinations of coherent states associated with the two degenerate states of the subsystem. The coefficients $\alpha$ and $\beta$ quantify the deviation of the state from that at critical coupling.

\subsection{Charging, storage and work extraction}
To design a suitable charged state, we looked at the closed Hamiltonian eigenvectors \eqref{eigen-closed} and construct an ad-hoc local unitary transformation (change of basis) acting on the qubits that could exploit a semi-DFS \cite{di2024optimal}. As discussed, the singlet state remains resilient to the influence of the bath, making it insensitive to decoherence. In what follows we will exploit then the semi-DFS when the system is in a state that is very close to $\ket{S}$ once charged.

\subsubsection{Discharged state}
We begin by determining the ground state $\ket{\psi_{GS}}$ (Eq.\,\eqref{gndstate}) of the entire Hamiltonian $H$, our discharged state, using the DMRG algorithm. We adiabatically apply a small magnetic field $H_{field}=-\epsilon (\sigma_z^1+\sigma_z^2)$ at $t=-\infty$, to favor the configuration $\ket{\uparrow\uparrow}$, as it will be one of the two degenerate states for increasing $g$, with $\alpha$ fixed. This approach allows us to move from the strong coupling regime to the ultrastrong coupling one. 
Once the transition occurs, the subsystem must to be only in the $\ket{\uparrow\uparrow}$ state.

We construct a transformation such that, in this case, the effect of the local operator is to bring the system into the singlet state alone. On the other hand, when $\beta$ is non-zero (for coupling $g$ below the critical point), the system can be written as a superposition of the two states (Eq.\,\eqref{gndstate}). In this scenario, after applying the local operator on the ground state, the resultant state is not an exact eigenstate of the entire Hamiltonian, and the interaction with the environment is not zero, leading to temporal evolution. This state can be characterized as a semi-DFS, as it is in a combination of the singlet state and the most excited one ($\ket{2}$ and $\ket{3}$ in \eqref{eigen-closed}). Only in the strongly ordered phase post-QPT the two-qubit system is in the exact DFS, protected from external perturbations, while the bath keeps evolving, described by a superposition of coherent states. However, in this case, the interaction with the system does not affect the bath dynamics.

\subsubsection{Charging}
Our protocol is based on the use of the DFS. Note that for this value of the two-qubit interaction strength $J$, thw DFS is very high in energy, allowing us to extract a significant amount of energy from it. The idea is to bring the state of the subsystem as close as possible to the DFS to protect the dynamics from the bath. 

We define then a local unitary operation $U_{S}^{(c)}$ that performs a change of basis from the computational basis $\left\{\ket{\uparrow\uparrow},\ket{\uparrow\downarrow},\ket{\downarrow\uparrow},\ket{\downarrow\downarrow}\right\}$ to the eigenstates basis \eqref{eigen-closed} of the closed Hamiltonian. The unitary matrix in the computational basis is
\begin{equation}
U_{S}^{(c)}=\frac{1}{\sqrt{2}}\begin{pmatrix}
    0 & -b & 1 & a \\
    1 & a & 0 & b\\
    -1 & a & 0 & b\\
    0 & -b & -1 & a \label{eq:ucharge}
\end{pmatrix},
\end{equation}
where $a$ and $b$ are the coefficients in the definition of the closed eigenstates \eqref{eigen-closed} in Eqs. \eqref{eq:afac} and \eqref{eq:bfac} (See Appendix \ref{app:unitary} for a circuit implementation). By applying this transformation on the computational basis, we get $U_{S}^{(c)}\ket{\uparrow\uparrow}= \ket{2};\,U_{S}^{(c)}\ket{\uparrow\downarrow}= \ket{0};U_{S}^{(c)}\ket{\downarrow\uparrow}= \ket{1};U_{S}^{(c)}\ket{\downarrow\downarrow}= \ket{3}$. This means that we are correctly mapping the two degenerate ground states to the most excited states of the closed system, both of which are very close to the singlet for our choice of large negative $J$.

With this, we are now prepared to compute the charged state
\begin{equation}
\label{charging}
    \ket{\psi^{(c)}}=(U_S^{(c)}\otimes\mathbb{1}_{E})\ket{\psi_{GS}}\,.
\end{equation}
Once we have charged the system, we can either immediately (say at time $t=0$) extract all the energy from the state by applying the inverse of the local operator $U_S^{(c)}$ on the qubits, or we can wait 
a certain time to extract work from the system (say at time $t>0$) testing the storage capabilities of the setup.
\subsubsection{Storage}
Once the state is charged, we may need to store the useful energy to make extraction possible at a subsequent time. At a generic time $t\geq0$, the full system is described by the state vector
\begin{equation}
    \label{evolution_psi}
\ket{\psi(t)}_{SE}=e^{-iH_{SE}t} \ket{\psi^{(c)}}\,,
\end{equation}
and the reduced density matrix is given by
\begin{equation}
    \label{evolution_rho}
    \rho_S(t)={\rm tr}_E(\ket{\psi(t)}\bra{\psi(t)}_{SE})\,.
\end{equation}
From the computational point of view, we allow the system to evolve with the entire Hamiltonian using TDVP numerical simulations. For both static and dynamic simulations we use ITensor library \cite{fishman2022itensor}. Following the application of the unitary operator in \eqref{eq:ucharge}, the system is in a semi-DFS state. Consequently, the two-qubit system remains in a quasi-singlet state throughout the evolution, maintaining its energy constant. However, the bath undergoes evolution, causing changes in the state of $SE$. This evolution is the underlying reason for the degradation of local ergotropy over time.  

\subsubsection{Local work extraction}
\paragraph*{Local work extraction at $t=0$.---}
In the absence of a storage phase ($t=0$), by applying the local unitary operator $U_S^{(c)}{}^\dag$ on $S$, the whole compound $SE$ goes back to the ground state, by only acting on the two-qubit subsystem. Then at time $t=0$,
by computing the local ergotropy $\mathcal{E}_{S}$, as in \eqref{local-ergo}, which coincides with the ergotropy $\mathcal{E}_{SE}$ of the $SE$ compound and with the excitation energy in this specific case, we have
\begin{equation}
\label{eq:ergotot}
\mathcal{E}_{S}=
    \mathcal{E}_{SE}=E_i-E_f=E_c-E_{GS},
    \quad \text{at $t=0,$}
\end{equation}
where the subscripts $i$ and $f$ signify the initial and final states, while $E_c=\bra{\psi_C}H\ket{\psi_C}$ is the energy of the charged state and $E_{GS}$ that of $\ket{\psi_{GS}}$. 

\paragraph*{Local work extraction at $t>0$.---}
Quantifying the local extractable work at a time $t>0$ serves to probe the quality of the storage phase aimed at conserving as much as possible the initial work resource.

We notice that in our model, 
at time $t>0$, as a consequence of the internal evolution of the compound $SE$,
we have 
\begin{equation}
\label{eq:storage}
\mathcal{E}_{S}(t)\leq 
    \mathcal{E}_{SE}=E_c-E_{GS},
    \quad \text{at $t>0\,.$}
\end{equation}
On this regard, we notice that in general, at variance with ergotropy \eqref{ergo}, an analytic expression for local ergotropy \eqref{local-ergo} is only known for the single qubit case \cite{LErgo}.
As a consequence, for $t>0$ the exact value of the local ergotropy $\mathcal{E}_S(t)$ is not known. However, given a generic unitary operation $U_S$ on $S$, the resulting work extraction 
\begin{equation}
\tilde{\mathcal{E}}_{S}:=
 \tr(H_{SE} \rho_{SE})-\tr(U_S \rho_{SE} U_S^\dag H_{SE})\,,
    \label{local-ergo-lb}
\end{equation}
will represent a lower bound for local ergotropy \eqref{local-ergo}.
Indeed, making explicit the dependence on $U_S$, we have 
\begin{equation}
\tilde{\mathcal{E}}_{S}(U_S)\leq 
{\mathcal{E}}_{S}\,,
\label{ineq}
\end{equation}
as ${\mathcal{E}}_{S}=\max_{U_S \in \mathcal{U}_{d_S}}\tilde{\mathcal{E}}_{S}(U_S)$. As discussed previously, in our protocol, the exact form of the two-qubit unitary operator allowing to saturate \eqref{ineq} is only known at time $t=0$. For times $t>0$ we then present two different approaches. 

Since the charged system is in a semi-DFS, we can try to extract work at time $t>0$ with the same unitary gate $U_S^{(c)}{}^\dag$ that was optimal at $t=0$. We call this \textit{agnostic} approach and the corresponding analysis is reported in Sec.\,\ref{sec:agnostic}.

The second approach, detailed in Sec.\,\ref{sec:optimization}, consists in what follows. For any given time $t>0$, we perform optimizations with different tools obtaining lower bounds $\tilde{\mathcal{E}}_{S}(t)$ that, if compared with the \textit{agnostic} protocol, are closer to the optimal value 
$\mathcal{E}_{S}(t)$.
\paragraph*{Work fluctuations.---}
We also compute the work fluctuations.
On one hand, this is to assess if they can be more sensitive to the presence of phase transitions when compared to the average work. On the other hand, from the technological point of view we want them to be not too large in order to obtain a suitable precision in work extraction. 

We notice that the charged state $\ket{\psi(t)}\bra{\psi(t)}_{SE}$, Eq.\,\eqref{charging}, does not commute with 
the compound's Hamiltonian $H_{SE}$. This implies that a two-point measurement (TPM) scheme \cite{esposito2009nonequilibrium} cannot describe the average value of the work extraction, as, e.g., provided by the local ergotropy when considering the maximization as in \eqref{local-ergo} \citep{perarnau-llobet_no-go_2017}. Indeed, it is well-known that the TPM scheme is invasive, in the sense that the first measurement of the energy destroys the initial coherence in the energy basis and this coherence generally represents a resource for work extraction. In this sense, quasiprobability distribution approaches are a useful tool to overcome this issue,
providing a way to describe the work average -- in our case, correctly yielding Eq.\,\eqref{local-ergo-lb} -- and variance in the presence of initial coherence \cite{francica2022class}.
The final expression for the work variance for our model reads as (see Ref.\,\cite{francica2022class} and Appendix \ref{app:variance}),
\begin{eqnarray}
&&
\sigma^2(t)=
\langle w^2\rangle-\langle w\rangle^2=
\\
&&
\tr[ (H' - H)^2 \rho_{SE}(t)]-
\{\tr[ (H' - H) \rho_{SE}(t)]\}^2
\,,\nonumber
\label{variance-t}
\end{eqnarray}
where $H':=U^\dag_S H U_S$, with $U_S$ being the chosen local unitary operation responsible for work extraction at time $t$, and $\rho_{SE}(t):=\ket{\psi(t)}\bra{\psi(t)}_{SE}$.

\section{Numerical Results}
\label{sec:results}
In this section, we showcase the results obtained for the charging process outlined in Sec.\,\ref{sec:charge}, as well as for the two methods employed to extract work from the system: the \textit{agnostic} approach using $U_S^{(c)}{}^{\dagger}$ detailed in Sec.\,\ref{sec:agnostic}, and the optimized method detailed in \ref{sec:optimization}. 

In the following, we set: $\omega_0=\Delta\,,J=-10\Delta\,,\alpha=0.1\,,\omega_c=30\Delta$ and take $\Delta$ as unit. Additionally, we employ DMRG to determine the ground state of $SE$ as an MPS with $N=300$ bath oscillators and a cutoff of $n=7$ for the Hilbert space of each oscillator. We ensure the convergence of the DMRG results by performing $10$ sweeps and controlling the minimum and maximum bond dimensions, the truncation error, the number of exact diagonalizations per sweep, and the noise term added to the density matrix. 

\subsection{Charging capability}\label{sec:charge}
\begin{figure}[htbp!]
    \begin{center}
        \includegraphics[scale=0.3]{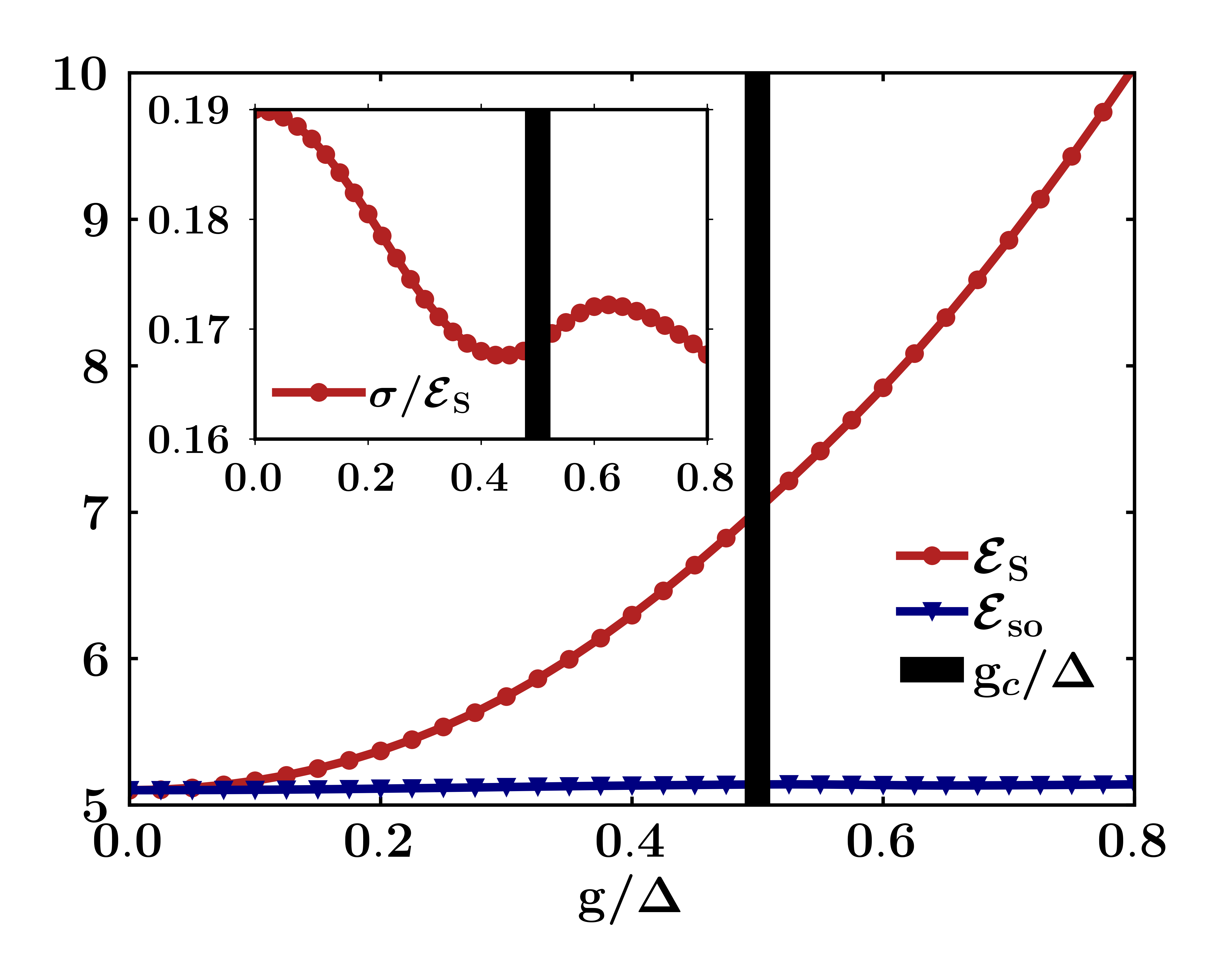}
     
        \caption{\label{fig:ergo_nostorage}
        Charging capability of the model:
        local (red circle) and switch-off (blue triangle) ergotropy in units of $\Delta$ for increasing $g/\Delta$, at time $t=0$, that is without a storage phase.In the inset we report the relative work fluctuations associated to the local ergotropy as a function of $g/\Delta$. We employ DMRG to determine the ground state of the $SE$.}
        
    \end{center}
\end{figure}

Here we present the analysis conducted for the charging process of the system defined in Section \ref{sec:model}. We compute local ergotropy (coinciding with the total ergotropy and energy excitation) along with its relative fluctuations and switch-off ergotropy for increasing values of $g$, as illustrated in Fig.\,\ref{fig:ergo_nostorage}.
The numerical results indicate that local ergotropy nearly doubles its switch-off counterpart (see Appendix \ref{app:so} for details on its computation) as $g$ increases, and it remains insensitive to the QPT occurring around $g\approx 0.5\Delta$. This firstly shows how the coupling to the environment is not detrimental to the work extraction, although it doesn't serve as a marker for the transition \cite{salvia2024zero}. This aligns with the BKT transition's infinite order nature, which implies non-analyticities in any derivative of complex system functions, rather than in energy. However, the growth of local ergotropy with $g$ implicitly depends on the QPT presence. The closer the system is to the critical $g_c$, the closer its state is to the singlet after applying the unitary gate. The inset of Fig.\,\ref{fig:ergo_nostorage} shows relative work fluctuations, namely the ratio between the work standard deviation $\sigma$ and the average extractable work $\mathcal{E}_S$, exhibiting a change in concavity, proving sensitivity to the QPT. Additionally, we highlight that the switch-off ergotropy value is dominated by the subsystem ergotropy contribution $\mathcal{E}_{sub}$ (Eq.\,\eqref{sub-ergo}). Furthermore, it is approximately on the order of the energy difference between the ground state and the most excited state of the closed system in Eq.\,\eqref{eigen-closed}, meaning we are extracting the maximum energy from the two-qubit system. Specifically, for very large negative $J$, such as in our case, this value is on the order of $-J/2\approx 5\Delta$. 

\subsection{Storage and work extraction features}
We allow the system to evolve with the entire Hamiltonian (see Eqs. \eqref{evolution_psi} and \eqref{evolution_rho}) and evaluate the work extraction capabilities as function of time.
\subsubsection{``Agnostic" approach for work extraction}
\label{sec:agnostic} 
\begin{figure}[htbp!]
    \begin{center}
        \includegraphics[scale=0.3]{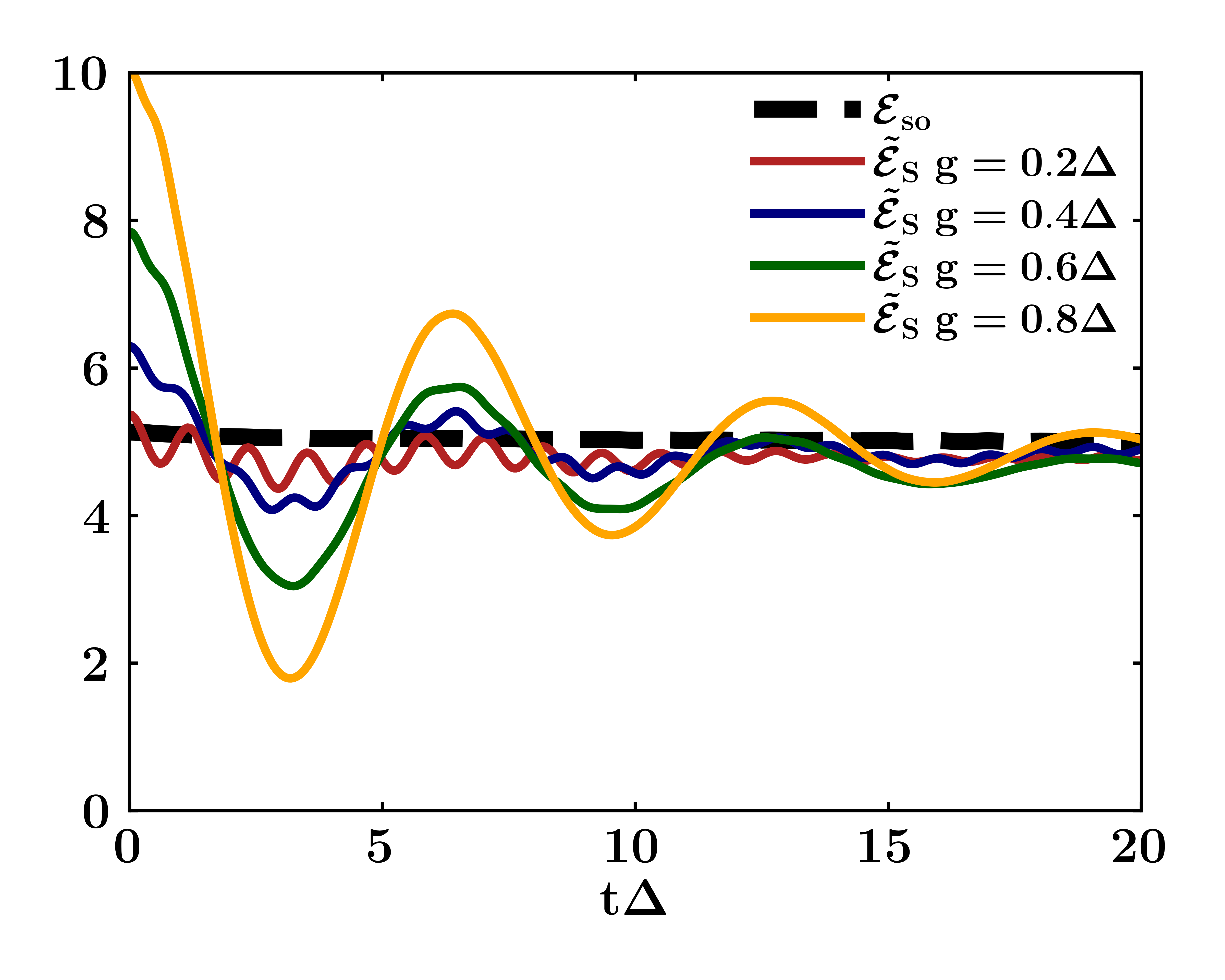}
     
        \caption{\label{fig:ergo_storage}Lower bound of local ergotropy (solid lines) in units of $\Delta$ obtained using the \textit{agnostic} protocol described in Sec.\,\ref{sec:agnostic}, 
        as function of dimensionless time, for increasing $g/\Delta$ (from red to orange). 
        We also report the switch-off ergotropy in units of $\Delta$ averaging over the almost constant values obtained for different $g/\Delta$ (black dashed line). We employ DMRG to determine the ground state of the $SE$. Then, we apply the unitary operation in \eqref{eq:ucharge} and evolve the charged state using TDVP. We extract work after applying the inverse of the local unitary gate \eqref{eq:ucharge} on the MPS.
        }
    \end{center}
\end{figure}

We initially present the results concerning the extractable work via the constant unitary gate $U_S^{(c)}{}^\dag$, namely the inverse of the charging unitary operator \eqref{eq:ucharge}, as a function of the storage time. In particular, we plot in Fig.\,\ref{fig:ergo_storage} the time evolution of the lower bound of the local ergotropy $\tilde{\mathcal{E}}(U_S^{(c)}{}^\dag)$ and the switch-off ergotropy ${\mathcal{E}}_{so}$ as functions of $t$ for increasing values of $g$, crossing the critical value $g_c\approx 0.5 \Delta$.
${\mathcal{E}}_{so}(t)$ is almost constant both in $t$ and in $g$ and, importantly, non-zero as a consequence of the fact that the charged state is quasi-decoherence free and $J$ having a large absolute value.

We observe that, as a consequence of a coherent interaction with the bath, the local ergotropy lower bound oscillates around the corresponding switch-off ergotropy, allowing one to obtain an improvement for long times. 

It's worth noting that the period $T$ of these oscillations depends on the parameters of the system and can be adjusted by changing $\omega_0$ since it is $T\approx 2\pi/\omega_0$ for couplings $g\geq g_c$. Additionally, the maxima and minima of the oscillations can be understood in terms of the in-phase and counter-phase behaviors of the bath oscillators. Our unitary gate takes the system in the semi-DFS ensuring that the two-qubit system does not interact with the bath, allowing the bath oscillators to evolve with their phase factors at frequencies near $\omega_0$. Consequently, the primary oscillation frequency is $\omega_0$. At time zero, there's a maximum due to constructive interference among the bath oscillators, whereas after half a period, they are in phase opposition, resulting in a minimum local ergotropy. Subsequently, at every $t=2k\pi/\omega_0$ with $k\in\mathbb{Z}$, a relative maximum occurs. The values of the maxima decrease over time due to the large number $N$ of harmonic oscillators. Once it becomes impossible to extract more energy from the two-qubit system using the bath energy, the local ergotropy remains at the switch-off level. It's important to note that it never reaches zero. 

Additionally, reducing the bath cutoff frequency $\omega_c$ to the system energy scale introduces non-Markovian effects, increasing the period, as it renormalizes the oscillator frequency filtering the frequencies above it in the bath spectrum. This feature can be utilized to engineer the protocol as needed.
Furthermore, we compute the work fluctuations concerning the \textit{agnostic} protocol, particularly interested in the relative fluctuations, again the ratio between the work standard deviation $\tilde{\sigma}$ and the average extractable work $\tilde{\mathcal{E}}(U_S^{(c)}{}^\dag)$, providing a quantification of the imprecision in the work extraction. 

\begin{figure}[htbp!]
    \begin{center}
        \includegraphics[scale=0.3]{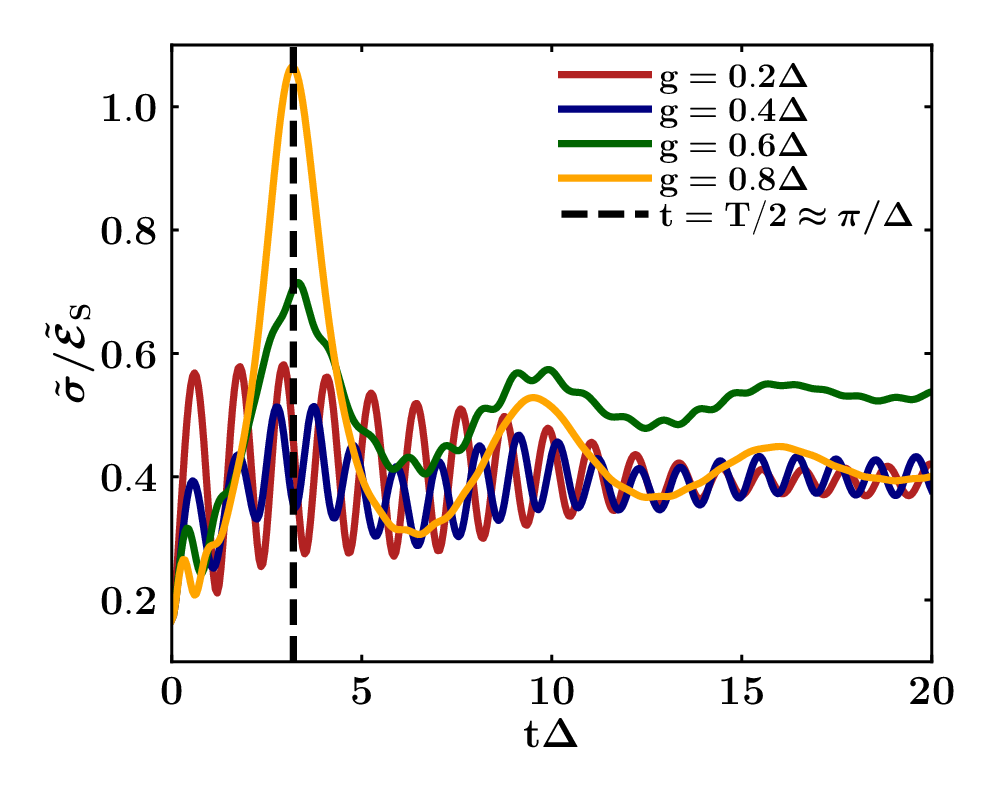}
     
        \caption{\label{fig:fluttergo_storage}Relative fluctuations of the lower bound of the local ergotropy (solid lines) as a function of dimensionless time for increasing $g/\Delta$ (from red to orange). The vertical dashed black line represents the time when the bath is in counterphase, farthest from the initial state. We employ DMRG to determine the ground state of the $SE$. Then, we apply the unitary operation in \eqref{eq:ucharge} and evolve the charged state using TDVP. We extract work after applying the local unitary gate \eqref{eq:ucharge} on the MPS.} 
    \end{center}
\end{figure}

In Fig.\,\ref{fig:fluttergo_storage}, we display $\tilde{\sigma}/\tilde{\mathcal{E}}(U_S^{(c)}{}^\dag)$ as a function of time for increasing values of $g$. 
A distinct behavior is observable between the low-$g$ curves and those beyond the critical $g$, attributed to the presence of the QPT and the effect of the counter-rotating terms in the Rabi model, becoming more significant as $g$ increases. 
Notably, a maximum emerges for high $g$ values, occurring when the bath is for the first time in counter-phase, farthest from its initial state. The maxima in the relative fluctuations behavior coincide with the bath being in counter-phase, corresponding to minima in local ergotropy.

To summarize, the change of the behavior with increasing $g$ can be observed in both local ergotropy and relative fluctuations, due to the presence of the QPT. For $g\geq g_c$, oscillations with $\omega_0$ emerge, and the minimum for the local ergotropy (maximum for relative fluctuations) at $t=\pi/\omega_0$ becomes increasingly prominent with higher $g$. This provides an alternative way to observe the occurrence of the QPT which, as an infinite order BKT transition, would be difficult to detect by examining typical system quantities.
\subsubsection{Optimized work extraction}
\label{sec:optimization}
In principle, to compute the local ergotropy \eqref{local-ergo} one should maximize the average local work extraction over all the possible two-qubit unitaries applied to the system. This is in general not an easy task. Therefore, we try to approximately compute the local ergotropy \eqref{local-ergo} or, more rigorously, a good lower bound for this quantity.
This should be at least greater than $\tilde{\mathcal{E}}(U_S^{(c)}{}^\dag)$. Hence, referring to Eq.\,\eqref{local-ergo-lb},
we parameterize the two-qubit unitary operation $U_{S}^{(e)}$ for work extraction through a suitable ansatz.
We take $U_{S}^{(e)}$ of the form
\begin{equation}
\label{ansatzU}
U_{S}^{(e)}(\theta,\phi):=U_{S}'(\theta,\phi)U_{S}^{(c)}{}^\dag\,,
\end{equation}
where 
\begin{equation}
U_{S}'(\theta,\phi):=
\begin{pmatrix}
e^{-i\phi}\sin({\theta}/{2})&0&0&\cos(\theta/{2})\\
0&1&0&0\\
0&0&1&0\\
-\cos(\theta/{2})&0&0&e^{i\phi}\sin({\theta}/{2})
\end{pmatrix}\,.
\label{rotation-subspace}
\end{equation}

The angles' ranges are taken as $\theta \in [0, \pi]$ and $\phi \in [0, 2\pi]$ and $a$ and $b$ are the same coefficients as those characterizing the closed eigenstates \eqref{eigen-closed} in Eqs. \eqref{eq:afac} and \eqref{eq:bfac} and the charging unitary gate \eqref{eq:ucharge}. We also have $U_{S}^{(e)}(\theta=\pi,\phi=0)=U_{S}^{(c)}{}^\dag$, implying that, if we optimize over $\theta$ and $\phi$, the unitary operator \eqref{ansatzU} will extract at least the work of $U_{S}^{(c)}{}^\dag$. The unitary matrix $U_{S}'(\theta,\phi)$ in Eq.\,\eqref{rotation-subspace} is the origin of the boost in performance. $U_{S}'(\theta,\phi)$ affects only the subspace spanned by the two ferromagnetic states, $\{\ket{\uparrow \uparrow},\ket{\downarrow \downarrow}\}$, that are the most probable near the QPT \cite{3-angles}. On a practical ground, it allows one to perform a rotation finalized to get closer to the correct superposition of $\ket{\uparrow \uparrow}$ and $\ket{\downarrow \downarrow}$ that characterizes the ground state \eqref{gndstate}, the target discharged state.

Hence, under the ansatz \eqref{ansatzU}, for any time $t>0$ we assess the maximum of the local extractable work (still representing a lower bound for local ergotropy) by varying $\theta$ and $\phi$ over an equally spaced grid, and picking the best values $(\bar{\theta}(t),\bar{\phi}(t))$. 

Finally, to test the quality of the local ergotropy lower bound based on the considered ansatz, we explore the optimum work extraction achievable by applying random Haar-distributed unitaries on the two-qubit system and maximizing the extractable work at any given time $t$. This allows us to quantify how far our results are from those obtained using random distributions.

\begin{figure}[htbp!]
    \begin{center}
        \includegraphics[scale=0.3]{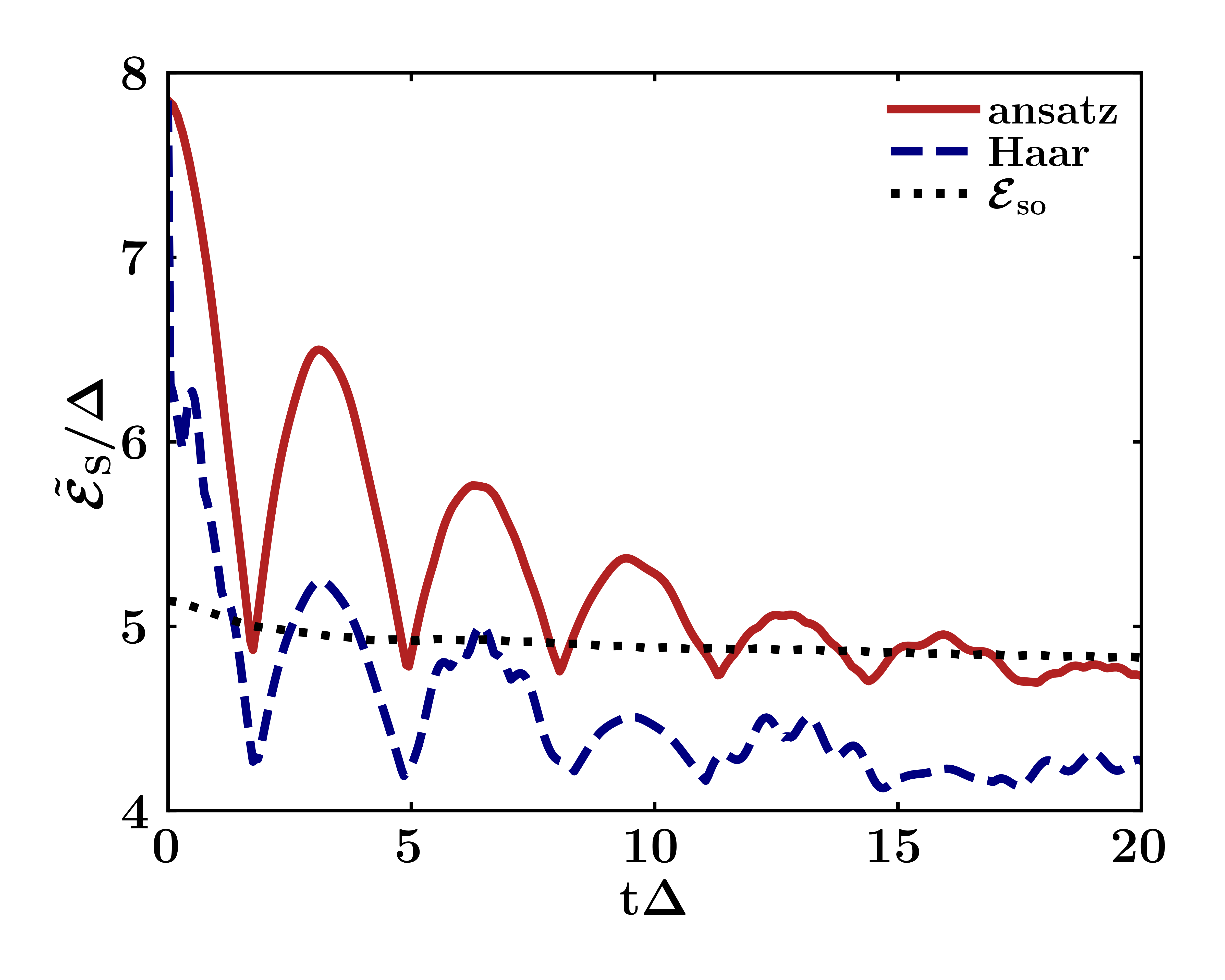}
     
        \caption{\label{fig:ergo_storage_opt}Local extractable work in units of $\Delta$ optimized through the ansatz \eqref{ansatzU} (red solid line) as a function of dimensionless time, for $g/\Delta=0.6$. The optimization is performed by varying $\theta$ and $\phi$ in \eqref{ansatzU} over an equally spaced grid. As in Fig.\,\ref{fig:ergo_storage}, this represents a lower bound for local ergotropy, but an improvement with respect to the green curve in Fig.\,\ref{fig:ergo_storage} corresponding to the same $g$ value. We also report the result obtained via an optimization drawing 100 random (converging over the average) Haar unitary operations (blue dashed line) and picking the one corresponding to the greatest local extractable work. The black dotted curve is the switch-off ergotropy. We employ DMRG to determine the ground state of the $SE$. Then, we apply the unitary operation in \eqref{eq:ucharge} and evolve the charged state using TDVP. We extract work after applying the local unitary gate \eqref{ansatzU} on the MPS for the ansatz and for the optimal parameters $(\bar\theta, \bar\phi)$ found for the Haar distribution.
        } 
    \end{center}
\end{figure}Figure \ref{fig:ergo_storage_opt} demonstrates that, once optimized over $\theta$ and $\phi$, the ansatz unitary gate \eqref{ansatzU} (red solid line) consistently outperforms the Haar unitaries (blue dashed line). This shows that our physically inspired ansatz lies among the outliers of a random Haar unitaries distribution. Knowing the phase diagram of this model helped us to find a simple and efficient ansatz for work extraction.

It is worth noting that we choose to show only the case for $g=0.6\Delta>g_c$, to ensure that the charging unitary gate mapped the ground state to the DFS. However, a similar behavior would be observed if considering $g\approx g_c$, but with reduced maxima and the addition of a superoscillation with a frequency related to the two-qubit energy $\Delta$.

\begin{figure}[htbp!]
    \begin{center}
        \includegraphics[scale=0.3]{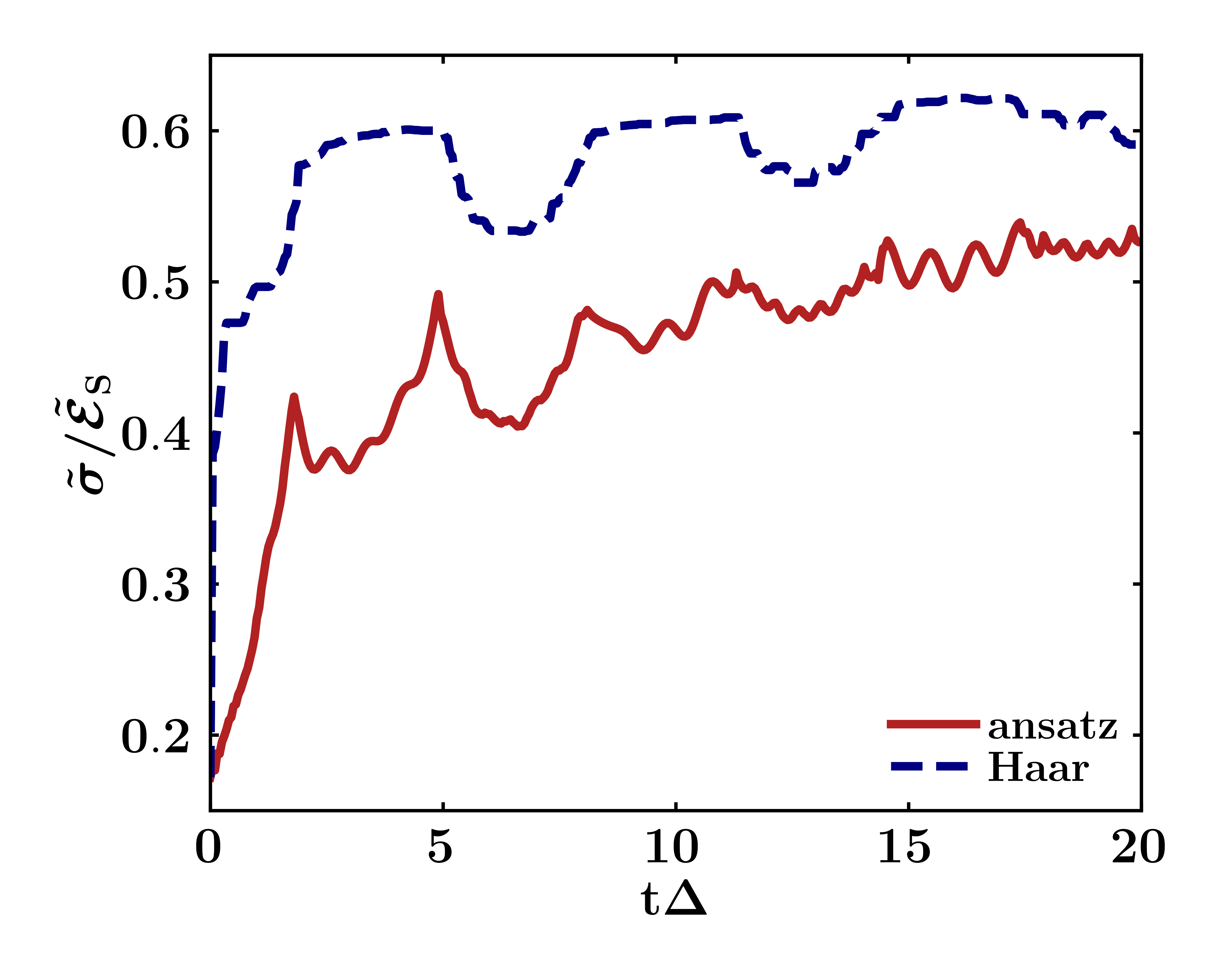}
     
        \caption{\label{fig:fluttergo_storage_opt}
        Relative work fluctuations associated to the lower bound of the local ergotropy reported in Fig.\,\ref{fig:ergo_storage_opt}, as a function of dimensionless time, for $g/\Delta=0.6$.
        They correspond to the optimized procedures described in Sec.\,\ref{sec:optimization} based on the ansatz \eqref{ansatzU} (red solid line) and drawing 100 random (converging over the average) Haar unitary operations (blue dashed line) and picking the one corresponding to the greatest local ergotropy. We employ DMRG to determine the ground state of the $SE$. Then, we apply the unitary operation in \eqref{eq:ucharge} and evolve the charged state using TDVP. We extract work after applying the local unitary gate \eqref{ansatzU} on the MPS for the ansatz and for the optimal parameters $(\bar\theta, \bar\phi)$ found for the Haar distribution.
        } 
    \end{center}
    \end{figure}

Interestingly, oscillations persist depending on the oscillator and the bath parameters, but the extractable work now, at variance with the \textit{agnostic} protocol, is always above or on the order of the switch-off ergotropy. This suggests that we are effectively rephasing the bath, transforming minima into maxima and thereby extending the time intervals during which we gain compared to the switch-off protocol. Furthermore, the ansatz approaches the switch-off ergotropy at long times (black dots). This indicates that over time, the advantage of interactions with the environment diminishes as it reaches equilibrium, far from the initial state, but the local ergotropy never reaches zero as well. We also compute the relative fluctuations for the lower bound of the local ergotropy (Fig.\,\ref{fig:fluttergo_storage_opt}), observing no transition markers but cutting the maximum as we approach the critical $g$. Put differently, this shows the optimized protocol to possess better stability in precision than the agnostic protocol. This is attributed to the presence of maxima instead of minima in the optimized lower bound of local ergotropy. Moreover, the fluctuations using random Haar matrices are always greater than the ones obtained through the physically inspired ansatz.

For completeness, in Appendix \ref{app:variability} we present average and standard deviation statistics for Haar unitaries compared with the ansatz, further confirming the efficacy of the latter, which stands at $2.5$ standard deviations from the Haar average.

Interestingly, in Appendix \ref{app:bayesian}, we describe another method we adopt to optimize the ergotropy. We use a Bayesian optimization approach over a unitary operator not very far from the optimized ansatz one, obtaining similar results with only a $1\%$ improvement. These results are thus not shown in the paper. 

\section{Conclusions}
\label{sec:conc}
We compute local ergotropy and its relative fluctuations in the two-qubit open Rabi model and implement a protocol for charging, storing, and discharging the two-qubit system, demonstrating its application as an open quantum battery. Our findings confirm that interaction with the external environment enhances, rather than hinders, work extraction. Specifically, increased coupling to an external bath doubles the local ergotropy immediately after system charging. Furthermore, when comparing local ergotropy with switch-off ergotropy, the former is approximately twice the latter. Indeed, the switch-off ergotropy considers the interaction with the environment only through the energy required to switch it off.

We observed that even though local ergotropy seems to be insensitive to the phase transition, always increasing with increasing $g$ (see Fig.\,\ref{fig:ergo_nostorage}). This increase is due to the ad-hoc construction of the local unitary operation for charging, tailored to the ground state of the system in the presence of the transition. Additionally, our model exploits the DFS, achieved through a local unitary operation during charging, providing protection from external interactions during storage.

Dynamic analysis revealed oscillatory behaviors in ergotropy and its fluctuations (see Figs.\,\ref{fig:ergo_storage} and \ref{fig:fluttergo_storage}), with frequencies tied to the two-qubit system and the cavity, allowing further tuning of the quantum battery. These oscillations, influenced by the phase transition, underscore the intricate link between phase transitions and local ergotropy. The changes in their behavior near the transition can serve as an alternative marker of it. Moreover, we prove that local ergotropy during long time intervals exceeds switch-off ergotropy by exploiting coherent interaction with the bath.

By optimizing the discharge process through a physically inspired ansatz (see Figs.\,\ref{fig:ergo_storage_opt} and \ref{fig:fluttergo_storage_opt}), we achieved superior local ergotropy compared to switch-off ergotropy over time, approaching the switch-off ergotropy for long times, but never reaching zero.

Finally, it is worth estimating the size of the time windows we have analyzed in this work to understand if they can be resolved experimentally with current technology. This relies on factors such as the energy of an ultrastrongly coupled flux superconducting qubit device, which typically operates at several gigahertz. For instance, with $20t\Delta\approx 25 ns$, or by explicitly accounting for relaxation rates observed in devices reaching high coupling regimes. These rates are often estimated using the Lindblad approach and should be comparable to the previously estimated values.

Our work demonstrates the connection between BKT QPT and local ergotropy, identifying quantities sensitive to such transitions. Moreover, we suggest a feasible, experimentally tunable protocol for work extraction. Future research could explore the application of our protocol to larger systems, investigating the impact of various environmental interactions on ergotropy and phase transitions to provide deeper insights into the practical implementation of open quantum batteries.

\bibliography{bib}

\begin{thebibliography}{46}%
\makeatletter
\providecommand \@ifxundefined [1]{%
 \@ifx{#1\undefined}
}%
\providecommand \@ifnum [1]{%
 \ifnum #1\expandafter \@firstoftwo
 \else \expandafter \@secondoftwo
 \fi
}%
\providecommand \@ifx [1]{%
 \ifx #1\expandafter \@firstoftwo
 \else \expandafter \@secondoftwo
 \fi
}%
\providecommand \natexlab [1]{#1}%
\providecommand \enquote  [1]{``#1''}%
\providecommand \bibnamefont  [1]{#1}%
\providecommand \bibfnamefont [1]{#1}%
\providecommand \citenamefont [1]{#1}%
\providecommand \href@noop [0]{\@secondoftwo}%
\providecommand \href [0]{\begingroup \@sanitize@url \@href}%
\providecommand \@href[1]{\@@startlink{#1}\@@href}%
\providecommand \@@href[1]{\endgroup#1\@@endlink}%
\providecommand \@sanitize@url [0]{\catcode `\\12\catcode `\$12\catcode `\&12\catcode `\#12\catcode `\^12\catcode `\_12\catcode `\%12\relax}%
\providecommand \@@startlink[1]{}%
\providecommand \@@endlink[0]{}%
\providecommand \url  [0]{\begingroup\@sanitize@url \@url }%
\providecommand \@url [1]{\endgroup\@href {#1}{\urlprefix }}%
\providecommand \urlprefix  [0]{URL }%
\providecommand \Eprint [0]{\href }%
\providecommand \doibase [0]{http://dx.doi.org/}%
\providecommand \selectlanguage [0]{\@gobble}%
\providecommand \bibinfo  [0]{\@secondoftwo}%
\providecommand \bibfield  [0]{\@secondoftwo}%
\providecommand \translation [1]{[#1]}%
\providecommand \BibitemOpen [0]{}%
\providecommand \bibitemStop [0]{}%
\providecommand \bibitemNoStop [0]{.\EOS\space}%
\providecommand \EOS [0]{\spacefactor3000\relax}%
\providecommand \BibitemShut  [1]{\csname bibitem#1\endcsname}%
\let\auto@bib@innerbib\@empty
\bibitem [{\citenamefont {Allahverdyan}\ \emph {et~al.}(2004)\citenamefont {Allahverdyan}, \citenamefont {Balian},\ and\ \citenamefont {Nieuwenhuizen}}]{allahverdyan2004maximal}%
  \BibitemOpen
  \bibfield  {author} {\bibinfo {author} {\bibfnamefont {A.~E.}\ \bibnamefont {Allahverdyan}}, \bibinfo {author} {\bibfnamefont {R.}~\bibnamefont {Balian}}, \ and\ \bibinfo {author} {\bibfnamefont {T.~M.}\ \bibnamefont {Nieuwenhuizen}},\ }\href@noop {} {\bibfield  {journal} {\bibinfo  {journal} {Europhys. Lett.}\ }\textbf {\bibinfo {volume} {67}},\ \bibinfo {pages} {565} (\bibinfo {year} {2004})}\BibitemShut {NoStop}%
\bibitem [{\citenamefont {Choquehuanca}\ \emph {et~al.}(2024)\citenamefont {Choquehuanca}, \citenamefont {Obando}, \citenamefont {de~Paula},\ and\ \citenamefont {Sarandy}}]{choquehuanca2024dynamics}%
  \BibitemOpen
  \bibfield  {author} {\bibinfo {author} {\bibfnamefont {J.}~\bibnamefont {Choquehuanca}}, \bibinfo {author} {\bibfnamefont {P.}~\bibnamefont {Obando}}, \bibinfo {author} {\bibfnamefont {F.}~\bibnamefont {de~Paula}}, \ and\ \bibinfo {author} {\bibfnamefont {M.}~\bibnamefont {Sarandy}},\ }\href@noop {} {\bibfield  {journal} {\bibinfo  {journal} {arXiv preprint arXiv:2403.04698}\ } (\bibinfo {year} {2024})}\BibitemShut {NoStop}%
\bibitem [{\citenamefont {Song}\ \emph {et~al.}(2024)\citenamefont {Song}, \citenamefont {Liu}, \citenamefont {Zhou}, \citenamefont {Yang},\ and\ \citenamefont {An}}]{song2024remote}%
  \BibitemOpen
  \bibfield  {author} {\bibinfo {author} {\bibfnamefont {W.-L.}\ \bibnamefont {Song}}, \bibinfo {author} {\bibfnamefont {H.-B.}\ \bibnamefont {Liu}}, \bibinfo {author} {\bibfnamefont {B.}~\bibnamefont {Zhou}}, \bibinfo {author} {\bibfnamefont {W.-L.}\ \bibnamefont {Yang}}, \ and\ \bibinfo {author} {\bibfnamefont {J.-H.}\ \bibnamefont {An}},\ }\href@noop {} {\bibfield  {journal} {\bibinfo  {journal} {Phys. Rev. Lett.}\ }\textbf {\bibinfo {volume} {132}},\ \bibinfo {pages} {090401} (\bibinfo {year} {2024})}\BibitemShut {NoStop}%
\bibitem [{\citenamefont {Yang}\ and\ \citenamefont {Dou}(2023)}]{yang2023resonator}%
  \BibitemOpen
  \bibfield  {author} {\bibinfo {author} {\bibfnamefont {F.-M.}\ \bibnamefont {Yang}}\ and\ \bibinfo {author} {\bibfnamefont {F.-Q.}\ \bibnamefont {Dou}},\ }\href@noop {} {\bibfield  {journal} {\bibinfo  {journal} {arXiv preprint arXiv:2312.11006}\ } (\bibinfo {year} {2023})}\BibitemShut {NoStop}%
\bibitem [{\citenamefont {Barra}(2015)}]{barra_thermodynamic_2015}%
  \BibitemOpen
  \bibfield  {author} {\bibinfo {author} {\bibfnamefont {F.}~\bibnamefont {Barra}},\ }\href {\doibase 10.1038/srep14873} {\bibfield  {journal} {\bibinfo  {journal} {Sci. Rep.}\ }\textbf {\bibinfo {volume} {5}},\ \bibinfo {pages} {14873} (\bibinfo {year} {2015})}\BibitemShut {NoStop}%
\bibitem [{\citenamefont {Ito}\ and\ \citenamefont {Miyadera}(2017)}]{ito_fundamental_2017}%
  \BibitemOpen
  \bibfield  {author} {\bibinfo {author} {\bibfnamefont {K.}~\bibnamefont {Ito}}\ and\ \bibinfo {author} {\bibfnamefont {T.}~\bibnamefont {Miyadera}},\ }\href {\doibase 10.48550/arXiv.1711.02322} {\enquote {\bibinfo {title} {Fundamental bound on the power of quantum machines},}\ } (\bibinfo {year} {2017}),\ \bibinfo {note} {arXiv:1711.02322 [cond-mat, physics:quant-ph]}\BibitemShut {NoStop}%
\bibitem [{\citenamefont {Andolina}\ \emph {et~al.}(2018)\citenamefont {Andolina}, \citenamefont {Farina}, \citenamefont {Mari}, \citenamefont {Pellegrini}, \citenamefont {Giovannetti},\ and\ \citenamefont {Polini}}]{andolina_charger-mediated_2018}%
  \BibitemOpen
  \bibfield  {author} {\bibinfo {author} {\bibfnamefont {G.~M.}\ \bibnamefont {Andolina}}, \bibinfo {author} {\bibfnamefont {D.}~\bibnamefont {Farina}}, \bibinfo {author} {\bibfnamefont {A.}~\bibnamefont {Mari}}, \bibinfo {author} {\bibfnamefont {V.}~\bibnamefont {Pellegrini}}, \bibinfo {author} {\bibfnamefont {V.}~\bibnamefont {Giovannetti}}, \ and\ \bibinfo {author} {\bibfnamefont {M.}~\bibnamefont {Polini}},\ }\href {\doibase 10.1103/PhysRevB.98.205423} {\bibfield  {journal} {\bibinfo  {journal} {Phys. Rev. B}\ }\textbf {\bibinfo {volume} {98}},\ \bibinfo {pages} {205423} (\bibinfo {year} {2018})}\BibitemShut {NoStop}%
\bibitem [{\citenamefont {Chiara}\ \emph {et~al.}(2018)\citenamefont {Chiara}, \citenamefont {Landi}, \citenamefont {Hewgill}, \citenamefont {Reid}, \citenamefont {Ferraro}, \citenamefont {Roncaglia},\ and\ \citenamefont {Antezza}}]{chiara_reconciliation_2018}%
  \BibitemOpen
  \bibfield  {author} {\bibinfo {author} {\bibfnamefont {G.~D.}\ \bibnamefont {Chiara}}, \bibinfo {author} {\bibfnamefont {G.}~\bibnamefont {Landi}}, \bibinfo {author} {\bibfnamefont {A.}~\bibnamefont {Hewgill}}, \bibinfo {author} {\bibfnamefont {B.}~\bibnamefont {Reid}}, \bibinfo {author} {\bibfnamefont {A.}~\bibnamefont {Ferraro}}, \bibinfo {author} {\bibfnamefont {A.~J.}\ \bibnamefont {Roncaglia}}, \ and\ \bibinfo {author} {\bibfnamefont {M.}~\bibnamefont {Antezza}},\ }\href {\doibase 10.1088/1367-2630/aaecee} {\bibfield  {journal} {\bibinfo  {journal} {New J. Phys.}\ }\textbf {\bibinfo {volume} {20}},\ \bibinfo {pages} {113024} (\bibinfo {year} {2018})}\BibitemShut {NoStop}%
\bibitem [{\citenamefont {Strasberg}\ \emph {et~al.}(2017)\citenamefont {Strasberg}, \citenamefont {Schaller}, \citenamefont {Brandes},\ and\ \citenamefont {Esposito}}]{strasberg_quantum_2017}%
  \BibitemOpen
  \bibfield  {author} {\bibinfo {author} {\bibfnamefont {P.}~\bibnamefont {Strasberg}}, \bibinfo {author} {\bibfnamefont {G.}~\bibnamefont {Schaller}}, \bibinfo {author} {\bibfnamefont {T.}~\bibnamefont {Brandes}}, \ and\ \bibinfo {author} {\bibfnamefont {M.}~\bibnamefont {Esposito}},\ }\href {\doibase 10.1103/PhysRevX.7.021003} {\bibfield  {journal} {\bibinfo  {journal} {Phys. Rev. X}\ }\textbf {\bibinfo {volume} {7}},\ \bibinfo {pages} {021003} (\bibinfo {year} {2017})}\BibitemShut {NoStop}%
\bibitem [{\citenamefont {Salvia}\ \emph {et~al.}(2023)\citenamefont {Salvia}, \citenamefont {De~Palma},\ and\ \citenamefont {Giovannetti}}]{LErgo}%
  \BibitemOpen
  \bibfield  {author} {\bibinfo {author} {\bibfnamefont {R.}~\bibnamefont {Salvia}}, \bibinfo {author} {\bibfnamefont {G.}~\bibnamefont {De~Palma}}, \ and\ \bibinfo {author} {\bibfnamefont {V.}~\bibnamefont {Giovannetti}},\ }\href@noop {} {\bibfield  {journal} {\bibinfo  {journal} {Phys. Rev. A}\ }\textbf {\bibinfo {volume} {107}},\ \bibinfo {pages} {012405} (\bibinfo {year} {2023})}\BibitemShut {NoStop}%
\bibitem [{\citenamefont {Castellano}\ \emph {et~al.}(2024)\citenamefont {Castellano}, \citenamefont {Farina}, \citenamefont {Giovannetti},\ and\ \citenamefont {Acin}}]{castellano2024extended}%
  \BibitemOpen
  \bibfield  {author} {\bibinfo {author} {\bibfnamefont {R.}~\bibnamefont {Castellano}}, \bibinfo {author} {\bibfnamefont {D.}~\bibnamefont {Farina}}, \bibinfo {author} {\bibfnamefont {V.}~\bibnamefont {Giovannetti}}, \ and\ \bibinfo {author} {\bibfnamefont {A.}~\bibnamefont {Acin}},\ }\href@noop {} {\bibfield  {journal} {\bibinfo  {journal} {arXiv preprint arXiv:2401.10996}\ } (\bibinfo {year} {2024})}\BibitemShut {NoStop}%
\bibitem [{\citenamefont {Farina}\ \emph {et~al.}(2019)\citenamefont {Farina}, \citenamefont {Andolina}, \citenamefont {Mari}, \citenamefont {Polini},\ and\ \citenamefont {Giovannetti}}]{PhysRevB.99.035421}%
  \BibitemOpen
  \bibfield  {author} {\bibinfo {author} {\bibfnamefont {D.}~\bibnamefont {Farina}}, \bibinfo {author} {\bibfnamefont {G.~M.}\ \bibnamefont {Andolina}}, \bibinfo {author} {\bibfnamefont {A.}~\bibnamefont {Mari}}, \bibinfo {author} {\bibfnamefont {M.}~\bibnamefont {Polini}}, \ and\ \bibinfo {author} {\bibfnamefont {V.}~\bibnamefont {Giovannetti}},\ }\href {\doibase 10.1103/PhysRevB.99.035421} {\bibfield  {journal} {\bibinfo  {journal} {Phys. Rev. B}\ }\textbf {\bibinfo {volume} {99}},\ \bibinfo {pages} {035421} (\bibinfo {year} {2019})}\BibitemShut {NoStop}%
\bibitem [{\citenamefont {Breuer}\ and\ \citenamefont {Petruccione}(2002)}]{breuer2002theory}%
  \BibitemOpen
  \bibfield  {author} {\bibinfo {author} {\bibfnamefont {H.-P.}\ \bibnamefont {Breuer}}\ and\ \bibinfo {author} {\bibfnamefont {F.}~\bibnamefont {Petruccione}},\ }\href@noop {} {\emph {\bibinfo {title} {The theory of open quantum systems}}}\ (\bibinfo  {publisher} {OUP Oxford},\ \bibinfo {year} {2002})\BibitemShut {NoStop}%
\bibitem [{\citenamefont {Rossini}\ \emph {et~al.}(2019)\citenamefont {Rossini}, \citenamefont {Andolina},\ and\ \citenamefont {Polini}}]{rossini2019many}%
  \BibitemOpen
  \bibfield  {author} {\bibinfo {author} {\bibfnamefont {D.}~\bibnamefont {Rossini}}, \bibinfo {author} {\bibfnamefont {G.~M.}\ \bibnamefont {Andolina}}, \ and\ \bibinfo {author} {\bibfnamefont {M.}~\bibnamefont {Polini}},\ }\href@noop {} {\bibfield  {journal} {\bibinfo  {journal} {Phys. Rev. B}\ }\textbf {\bibinfo {volume} {100}},\ \bibinfo {pages} {115142} (\bibinfo {year} {2019})}\BibitemShut {NoStop}%
\bibitem [{\citenamefont {Rossini}\ \emph {et~al.}(2020)\citenamefont {Rossini}, \citenamefont {Andolina}, \citenamefont {Rosa}, \citenamefont {Carrega},\ and\ \citenamefont {Polini}}]{rossini2020quantum}%
  \BibitemOpen
  \bibfield  {author} {\bibinfo {author} {\bibfnamefont {D.}~\bibnamefont {Rossini}}, \bibinfo {author} {\bibfnamefont {G.~M.}\ \bibnamefont {Andolina}}, \bibinfo {author} {\bibfnamefont {D.}~\bibnamefont {Rosa}}, \bibinfo {author} {\bibfnamefont {M.}~\bibnamefont {Carrega}}, \ and\ \bibinfo {author} {\bibfnamefont {M.}~\bibnamefont {Polini}},\ }\href@noop {} {\bibfield  {journal} {\bibinfo  {journal} {Phys. Rev. Lett.}\ }\textbf {\bibinfo {volume} {125}},\ \bibinfo {pages} {236402} (\bibinfo {year} {2020})}\BibitemShut {NoStop}%
\bibitem [{\citenamefont {Zhao}\ \emph {et~al.}(2021)\citenamefont {Zhao}, \citenamefont {Dou},\ and\ \citenamefont {Zhao}}]{zhao2021quantum}%
  \BibitemOpen
  \bibfield  {author} {\bibinfo {author} {\bibfnamefont {F.}~\bibnamefont {Zhao}}, \bibinfo {author} {\bibfnamefont {F.-Q.}\ \bibnamefont {Dou}}, \ and\ \bibinfo {author} {\bibfnamefont {Q.}~\bibnamefont {Zhao}},\ }\href@noop {} {\bibfield  {journal} {\bibinfo  {journal} {Phys. Rev. A}\ }\textbf {\bibinfo {volume} {103}},\ \bibinfo {pages} {033715} (\bibinfo {year} {2021})}\BibitemShut {NoStop}%
\bibitem [{\citenamefont {Zhao}\ \emph {et~al.}(2022)\citenamefont {Zhao}, \citenamefont {Dou},\ and\ \citenamefont {Zhao}}]{zhao2022charging}%
  \BibitemOpen
  \bibfield  {author} {\bibinfo {author} {\bibfnamefont {F.}~\bibnamefont {Zhao}}, \bibinfo {author} {\bibfnamefont {F.-Q.}\ \bibnamefont {Dou}}, \ and\ \bibinfo {author} {\bibfnamefont {Q.}~\bibnamefont {Zhao}},\ }\href@noop {} {\bibfield  {journal} {\bibinfo  {journal} {Phys. Rev. Res.}\ }\textbf {\bibinfo {volume} {4}},\ \bibinfo {pages} {013172} (\bibinfo {year} {2022})}\BibitemShut {NoStop}%
\bibitem [{\citenamefont {Barra}\ \emph {et~al.}(2022)\citenamefont {Barra}, \citenamefont {Hovhannisyan},\ and\ \citenamefont {Imparato}}]{barra2022quantum}%
  \BibitemOpen
  \bibfield  {author} {\bibinfo {author} {\bibfnamefont {F.}~\bibnamefont {Barra}}, \bibinfo {author} {\bibfnamefont {K.~V.}\ \bibnamefont {Hovhannisyan}}, \ and\ \bibinfo {author} {\bibfnamefont {A.}~\bibnamefont {Imparato}},\ }\href@noop {} {\bibfield  {journal} {\bibinfo  {journal} {New J. Phys.}\ }\textbf {\bibinfo {volume} {24}},\ \bibinfo {pages} {015003} (\bibinfo {year} {2022})}\BibitemShut {NoStop}%
\bibitem [{\citenamefont {Arjmandi}\ \emph {et~al.}(2023)\citenamefont {Arjmandi}, \citenamefont {Mohammadi}, \citenamefont {Saguia}, \citenamefont {Sarandy},\ and\ \citenamefont {Santos}}]{arjmandi2023localization}%
  \BibitemOpen
  \bibfield  {author} {\bibinfo {author} {\bibfnamefont {M.~B.}\ \bibnamefont {Arjmandi}}, \bibinfo {author} {\bibfnamefont {H.}~\bibnamefont {Mohammadi}}, \bibinfo {author} {\bibfnamefont {A.}~\bibnamefont {Saguia}}, \bibinfo {author} {\bibfnamefont {M.~S.}\ \bibnamefont {Sarandy}}, \ and\ \bibinfo {author} {\bibfnamefont {A.~C.}\ \bibnamefont {Santos}},\ }\href@noop {} {\bibfield  {journal} {\bibinfo  {journal} {Phys. Rev. E}\ }\textbf {\bibinfo {volume} {108}},\ \bibinfo {pages} {064106} (\bibinfo {year} {2023})}\BibitemShut {NoStop}%
\bibitem [{\citenamefont {Grazi}\ \emph {et~al.}(2024)\citenamefont {Grazi}, \citenamefont {Shaikh}, \citenamefont {Sassetti}, \citenamefont {Ziani},\ and\ \citenamefont {Ferraro}}]{grazi2024enhancing}%
  \BibitemOpen
  \bibfield  {author} {\bibinfo {author} {\bibfnamefont {R.}~\bibnamefont {Grazi}}, \bibinfo {author} {\bibfnamefont {D.~S.}\ \bibnamefont {Shaikh}}, \bibinfo {author} {\bibfnamefont {M.}~\bibnamefont {Sassetti}}, \bibinfo {author} {\bibfnamefont {N.~T.}\ \bibnamefont {Ziani}}, \ and\ \bibinfo {author} {\bibfnamefont {D.}~\bibnamefont {Ferraro}},\ }\href@noop {} {\bibfield  {journal} {\bibinfo  {journal} {arXiv preprint arXiv:2402.09169}\ } (\bibinfo {year} {2024})}\BibitemShut {NoStop}%
\bibitem [{\citenamefont {Feli{\'u}}\ and\ \citenamefont {Barra}(2024)}]{feliu2024system}%
  \BibitemOpen
  \bibfield  {author} {\bibinfo {author} {\bibfnamefont {D.}~\bibnamefont {Feli{\'u}}}\ and\ \bibinfo {author} {\bibfnamefont {F.}~\bibnamefont {Barra}},\ }\href@noop {} {\bibfield  {journal} {\bibinfo  {journal} {arXiv preprint arXiv:2403.08573}\ } (\bibinfo {year} {2024})}\BibitemShut {NoStop}%
\bibitem [{\citenamefont {Gyhm}\ and\ \citenamefont {Fischer}(2024)}]{gyhm2024beneficial}%
  \BibitemOpen
  \bibfield  {author} {\bibinfo {author} {\bibfnamefont {J.-Y.}\ \bibnamefont {Gyhm}}\ and\ \bibinfo {author} {\bibfnamefont {U.~R.}\ \bibnamefont {Fischer}},\ }\href@noop {} {\bibfield  {journal} {\bibinfo  {journal} {AVS Quantum Science}\ }\textbf {\bibinfo {volume} {6}} (\bibinfo {year} {2024})}\BibitemShut {NoStop}%
\bibitem [{\citenamefont {Campaioli}\ \emph {et~al.}(2024)\citenamefont {Campaioli}, \citenamefont {Gherardini}, \citenamefont {Quach}, \citenamefont {Polini},\ and\ \citenamefont {Andolina}}]{RevModPhys.96.031001}%
  \BibitemOpen
  \bibfield  {author} {\bibinfo {author} {\bibfnamefont {F.}~\bibnamefont {Campaioli}}, \bibinfo {author} {\bibfnamefont {S.}~\bibnamefont {Gherardini}}, \bibinfo {author} {\bibfnamefont {J.~Q.}\ \bibnamefont {Quach}}, \bibinfo {author} {\bibfnamefont {M.}~\bibnamefont {Polini}}, \ and\ \bibinfo {author} {\bibfnamefont {G.~M.}\ \bibnamefont {Andolina}},\ }\href {\doibase 10.1103/RevModPhys.96.031001} {\bibfield  {journal} {\bibinfo  {journal} {Rev. Mod. Phys.}\ }\textbf {\bibinfo {volume} {96}},\ \bibinfo {pages} {031001} (\bibinfo {year} {2024})}\BibitemShut {NoStop}%
\bibitem [{\citenamefont {Niu}(2024)}]{niu2024work}%
  \BibitemOpen
  \bibfield  {author} {\bibinfo {author} {\bibfnamefont {Z.-X.}\ \bibnamefont {Niu}},\ }\href@noop {} {\bibfield  {journal} {\bibinfo  {journal} {arXiv preprint arXiv:2401.05921}\ } (\bibinfo {year} {2024})}\BibitemShut {NoStop}%
\bibitem [{\citenamefont {Liu}\ \emph {et~al.}(2019)\citenamefont {Liu}, \citenamefont {Segal},\ and\ \citenamefont {Hanna}}]{liu2019loss}%
  \BibitemOpen
  \bibfield  {author} {\bibinfo {author} {\bibfnamefont {J.}~\bibnamefont {Liu}}, \bibinfo {author} {\bibfnamefont {D.}~\bibnamefont {Segal}}, \ and\ \bibinfo {author} {\bibfnamefont {G.}~\bibnamefont {Hanna}},\ }\href@noop {} {\bibfield  {journal} {\bibinfo  {journal} {J. Phys. Chem. C}\ }\textbf {\bibinfo {volume} {123}},\ \bibinfo {pages} {18303} (\bibinfo {year} {2019})}\BibitemShut {NoStop}%
\bibitem [{\citenamefont {Crescente}\ \emph {et~al.}(2024)\citenamefont {Crescente}, \citenamefont {Ferraro},\ and\ \citenamefont {Sassetti}}]{crescente2024boosting}%
  \BibitemOpen
  \bibfield  {author} {\bibinfo {author} {\bibfnamefont {A.}~\bibnamefont {Crescente}}, \bibinfo {author} {\bibfnamefont {D.}~\bibnamefont {Ferraro}}, \ and\ \bibinfo {author} {\bibfnamefont {M.}~\bibnamefont {Sassetti}},\ }\href@noop {} {\bibfield  {journal} {\bibinfo  {journal} {Phys. Rev. Res.}\ }\textbf {\bibinfo {volume} {6}},\ \bibinfo {pages} {023092} (\bibinfo {year} {2024})}\BibitemShut {NoStop}%
\bibitem [{\citenamefont {White}(1992)}]{white_92}%
  \BibitemOpen
  \bibfield  {author} {\bibinfo {author} {\bibfnamefont {S.~R.}\ \bibnamefont {White}},\ }\href {\doibase 10.1103/PhysRevLett.69.2863} {\bibfield  {journal} {\bibinfo  {journal} {Phys. Rev. Lett.}\ }\textbf {\bibinfo {volume} {69}},\ \bibinfo {pages} {2863} (\bibinfo {year} {1992})}\BibitemShut {NoStop}%
\bibitem [{\citenamefont {Schollwöck}(2011)}]{schollwoeck_2011}%
  \BibitemOpen
  \bibfield  {author} {\bibinfo {author} {\bibfnamefont {U.}~\bibnamefont {Schollwöck}},\ }\href {\doibase https://doi.org/10.1016/j.aop.2010.09.012} {\bibfield  {journal} {\bibinfo  {journal} {Ann. Phys.}\ }\textbf {\bibinfo {volume} {326}},\ \bibinfo {pages} {96} (\bibinfo {year} {2011})},\ \bibinfo {note} {january 2011 Special Issue}\BibitemShut {NoStop}%
\bibitem [{\citenamefont {Haegeman}\ \emph {et~al.}(2011)\citenamefont {Haegeman}, \citenamefont {Cirac}, \citenamefont {Osborne}, \citenamefont {Pi\ifmmode~\check{z}\else \v{z}\fi{}orn}, \citenamefont {Verschelde},\ and\ \citenamefont {Verstraete}}]{haegeman_11}%
  \BibitemOpen
  \bibfield  {author} {\bibinfo {author} {\bibfnamefont {J.}~\bibnamefont {Haegeman}}, \bibinfo {author} {\bibfnamefont {J.~I.}\ \bibnamefont {Cirac}}, \bibinfo {author} {\bibfnamefont {T.~J.}\ \bibnamefont {Osborne}}, \bibinfo {author} {\bibfnamefont {I.}~\bibnamefont {Pi\ifmmode~\check{z}\else \v{z}\fi{}orn}}, \bibinfo {author} {\bibfnamefont {H.}~\bibnamefont {Verschelde}}, \ and\ \bibinfo {author} {\bibfnamefont {F.}~\bibnamefont {Verstraete}},\ }\href {\doibase 10.1103/PhysRevLett.107.070601} {\bibfield  {journal} {\bibinfo  {journal} {Phys. Rev. Lett.}\ }\textbf {\bibinfo {volume} {107}},\ \bibinfo {pages} {070601} (\bibinfo {year} {2011})}\BibitemShut {NoStop}%
\bibitem [{\citenamefont {Haegeman}\ \emph {et~al.}(2016)\citenamefont {Haegeman}, \citenamefont {Lubich}, \citenamefont {Oseledets}, \citenamefont {Vandereycken},\ and\ \citenamefont {Verstraete}}]{haegeman_16}%
  \BibitemOpen
  \bibfield  {author} {\bibinfo {author} {\bibfnamefont {J.}~\bibnamefont {Haegeman}}, \bibinfo {author} {\bibfnamefont {C.}~\bibnamefont {Lubich}}, \bibinfo {author} {\bibfnamefont {I.}~\bibnamefont {Oseledets}}, \bibinfo {author} {\bibfnamefont {B.}~\bibnamefont {Vandereycken}}, \ and\ \bibinfo {author} {\bibfnamefont {F.}~\bibnamefont {Verstraete}},\ }\href {\doibase 10.1103/PhysRevB.94.165116} {\bibfield  {journal} {\bibinfo  {journal} {Phys. Rev. B}\ }\textbf {\bibinfo {volume} {94}},\ \bibinfo {pages} {165116} (\bibinfo {year} {2016})}\BibitemShut {NoStop}%
\bibitem [{\citenamefont {Di~Bello}\ \emph {et~al.}(2024)\citenamefont {Di~Bello}, \citenamefont {De~Filippis}, \citenamefont {Hamma},\ and\ \citenamefont {Perroni}}]{di2024optimal}%
  \BibitemOpen
  \bibfield  {author} {\bibinfo {author} {\bibfnamefont {G.}~\bibnamefont {Di~Bello}}, \bibinfo {author} {\bibfnamefont {G.}~\bibnamefont {De~Filippis}}, \bibinfo {author} {\bibfnamefont {A.}~\bibnamefont {Hamma}}, \ and\ \bibinfo {author} {\bibfnamefont {C.~A.}\ \bibnamefont {Perroni}},\ }\href@noop {} {\bibfield  {journal} {\bibinfo  {journal} {Phys. Rev. B}\ }\textbf {\bibinfo {volume} {109}},\ \bibinfo {pages} {014304} (\bibinfo {year} {2024})}\BibitemShut {NoStop}%
\bibitem [{\citenamefont {Di~Bello}\ \emph {et~al.}(2023)\citenamefont {Di~Bello}, \citenamefont {Ponticelli}, \citenamefont {Pavan}, \citenamefont {Cataudella}, \citenamefont {De~Filippis}, \citenamefont {de~Candia},\ and\ \citenamefont {Perroni}}]{di2023environment}%
  \BibitemOpen
  \bibfield  {author} {\bibinfo {author} {\bibfnamefont {G.}~\bibnamefont {Di~Bello}}, \bibinfo {author} {\bibfnamefont {A.}~\bibnamefont {Ponticelli}}, \bibinfo {author} {\bibfnamefont {F.}~\bibnamefont {Pavan}}, \bibinfo {author} {\bibfnamefont {V.}~\bibnamefont {Cataudella}}, \bibinfo {author} {\bibfnamefont {G.}~\bibnamefont {De~Filippis}}, \bibinfo {author} {\bibfnamefont {A.}~\bibnamefont {de~Candia}}, \ and\ \bibinfo {author} {\bibfnamefont {C.~A.}\ \bibnamefont {Perroni}},\ }\href@noop {} {\bibfield  {journal} {\bibinfo  {journal} {arXiv preprint arXiv:2312.05697}\ } (\bibinfo {year} {2023})}\BibitemShut {NoStop}%
\bibitem [{\citenamefont {Caldeira}\ and\ \citenamefont {Leggett}(1981)}]{CaldeiraLeggett81}%
  \BibitemOpen
  \bibfield  {author} {\bibinfo {author} {\bibfnamefont {A.~O.}\ \bibnamefont {Caldeira}}\ and\ \bibinfo {author} {\bibfnamefont {A.~J.}\ \bibnamefont {Leggett}},\ }\href {\doibase 10.1103/PhysRevLett.46.211} {\bibfield  {journal} {\bibinfo  {journal} {Phys. Rev. Lett.}\ }\textbf {\bibinfo {volume} {46}},\ \bibinfo {pages} {211} (\bibinfo {year} {1981})}\BibitemShut {NoStop}%
\bibitem [{\citenamefont {Weiss}(2012)}]{weiss2012quantum}%
  \BibitemOpen
  \bibfield  {author} {\bibinfo {author} {\bibfnamefont {U.}~\bibnamefont {Weiss}},\ }\href@noop {} {\emph {\bibinfo {title} {Quantum dissipative systems}}}\ (\bibinfo  {publisher} {World Scientific},\ \bibinfo {year} {2012})\BibitemShut {NoStop}%
\bibitem [{\citenamefont {Magazz{\`u}}\ and\ \citenamefont {Grifoni}(2019)}]{magazzu2019transmission}%
  \BibitemOpen
  \bibfield  {author} {\bibinfo {author} {\bibfnamefont {L.}~\bibnamefont {Magazz{\`u}}}\ and\ \bibinfo {author} {\bibfnamefont {M.}~\bibnamefont {Grifoni}},\ }\href@noop {} {\bibfield  {journal} {\bibinfo  {journal} {J. Stat. Mech. Theory Exp.}\ }\textbf {\bibinfo {volume} {2019}},\ \bibinfo {pages} {104002} (\bibinfo {year} {2019})}\BibitemShut {NoStop}%
\bibitem [{\citenamefont {Goorden}\ \emph {et~al.}(2004)\citenamefont {Goorden}, \citenamefont {Thorwart},\ and\ \citenamefont {Grifoni}}]{goorden2004entanglement}%
  \BibitemOpen
  \bibfield  {author} {\bibinfo {author} {\bibfnamefont {M.}~\bibnamefont {Goorden}}, \bibinfo {author} {\bibfnamefont {M.}~\bibnamefont {Thorwart}}, \ and\ \bibinfo {author} {\bibfnamefont {M.}~\bibnamefont {Grifoni}},\ }\href@noop {} {\bibfield  {journal} {\bibinfo  {journal} {Phys. Rev. Lett.}\ }\textbf {\bibinfo {volume} {93}},\ \bibinfo {pages} {267005} (\bibinfo {year} {2004})}\BibitemShut {NoStop}%
\bibitem [{\citenamefont {De~Filippis}\ \emph {et~al.}(2023)\citenamefont {De~Filippis}, \citenamefont {de~Candia}, \citenamefont {Di~Bello}, \citenamefont {Perroni}, \citenamefont {Cangemi}, \citenamefont {Nocera}, \citenamefont {Sassetti}, \citenamefont {Fazio},\ and\ \citenamefont {Cataudella}}]{de2023signatures}%
  \BibitemOpen
  \bibfield  {author} {\bibinfo {author} {\bibfnamefont {G.}~\bibnamefont {De~Filippis}}, \bibinfo {author} {\bibfnamefont {A.}~\bibnamefont {de~Candia}}, \bibinfo {author} {\bibfnamefont {G.}~\bibnamefont {Di~Bello}}, \bibinfo {author} {\bibfnamefont {C.~A.}\ \bibnamefont {Perroni}}, \bibinfo {author} {\bibfnamefont {L.~M.}\ \bibnamefont {Cangemi}}, \bibinfo {author} {\bibfnamefont {A.}~\bibnamefont {Nocera}}, \bibinfo {author} {\bibfnamefont {M.}~\bibnamefont {Sassetti}}, \bibinfo {author} {\bibfnamefont {R.}~\bibnamefont {Fazio}}, \ and\ \bibinfo {author} {\bibfnamefont {V.}~\bibnamefont {Cataudella}},\ }\href@noop {} {\bibfield  {journal} {\bibinfo  {journal} {Phys. Rev. Lett.}\ }\textbf {\bibinfo {volume} {130}},\ \bibinfo {pages} {210404} (\bibinfo {year} {2023})}\BibitemShut {NoStop}%
\bibitem [{\citenamefont {Heyl}(2018)}]{heyl2018dynamical}%
  \BibitemOpen
  \bibfield  {author} {\bibinfo {author} {\bibfnamefont {M.}~\bibnamefont {Heyl}},\ }\href@noop {} {\bibfield  {journal} {\bibinfo  {journal} {Rep. Prog. Phys.}\ }\textbf {\bibinfo {volume} {81}},\ \bibinfo {pages} {054001} (\bibinfo {year} {2018})}\BibitemShut {NoStop}%
\bibitem [{\citenamefont {Fishman}\ \emph {et~al.}(2022)\citenamefont {Fishman}, \citenamefont {White},\ and\ \citenamefont {Stoudenmire}}]{fishman2022itensor}%
  \BibitemOpen
  \bibfield  {author} {\bibinfo {author} {\bibfnamefont {M.}~\bibnamefont {Fishman}}, \bibinfo {author} {\bibfnamefont {S.}~\bibnamefont {White}}, \ and\ \bibinfo {author} {\bibfnamefont {E.}~\bibnamefont {Stoudenmire}},\ }\href@noop {} {\bibfield  {journal} {\bibinfo  {journal} {SciPost Physics Codebases}\ ,\ \bibinfo {pages} {004}} (\bibinfo {year} {2022})}\BibitemShut {NoStop}%
\bibitem [{\citenamefont {Esposito}\ \emph {et~al.}(2009)\citenamefont {Esposito}, \citenamefont {Harbola},\ and\ \citenamefont {Mukamel}}]{esposito2009nonequilibrium}%
  \BibitemOpen
  \bibfield  {author} {\bibinfo {author} {\bibfnamefont {M.}~\bibnamefont {Esposito}}, \bibinfo {author} {\bibfnamefont {U.}~\bibnamefont {Harbola}}, \ and\ \bibinfo {author} {\bibfnamefont {S.}~\bibnamefont {Mukamel}},\ }\href@noop {} {\bibfield  {journal} {\bibinfo  {journal} {Rev. Mod. Phys.}\ }\textbf {\bibinfo {volume} {81}},\ \bibinfo {pages} {1665} (\bibinfo {year} {2009})}\BibitemShut {NoStop}%
\bibitem [{\citenamefont {Perarnau-Llobet}\ \emph {et~al.}(2017)\citenamefont {Perarnau-Llobet}, \citenamefont {Bäumer}, \citenamefont {Hovhannisyan}, \citenamefont {Huber},\ and\ \citenamefont {Acin}}]{perarnau-llobet_no-go_2017}%
  \BibitemOpen
  \bibfield  {author} {\bibinfo {author} {\bibfnamefont {M.}~\bibnamefont {Perarnau-Llobet}}, \bibinfo {author} {\bibfnamefont {E.}~\bibnamefont {Bäumer}}, \bibinfo {author} {\bibfnamefont {K.~V.}\ \bibnamefont {Hovhannisyan}}, \bibinfo {author} {\bibfnamefont {M.}~\bibnamefont {Huber}}, \ and\ \bibinfo {author} {\bibfnamefont {A.}~\bibnamefont {Acin}},\ }\href {\doibase 10.1103/PhysRevLett.118.070601} {\bibfield  {journal} {\bibinfo  {journal} {Phys. Rev. Lett.}\ }\textbf {\bibinfo {volume} {118}},\ \bibinfo {pages} {070601} (\bibinfo {year} {2017})}\BibitemShut {NoStop}%
\bibitem [{\citenamefont {Francica}(2022)}]{francica2022class}%
  \BibitemOpen
  \bibfield  {author} {\bibinfo {author} {\bibfnamefont {G.}~\bibnamefont {Francica}},\ }\href@noop {} {\bibfield  {journal} {\bibinfo  {journal} {Phys. Rev. E}\ }\textbf {\bibinfo {volume} {105}},\ \bibinfo {pages} {014101} (\bibinfo {year} {2022})}\BibitemShut {NoStop}%
\bibitem [{\citenamefont {Salvia}\ and\ \citenamefont {Giovannetti}(2024)}]{salvia2024zero}%
  \BibitemOpen
  \bibfield  {author} {\bibinfo {author} {\bibfnamefont {R.}~\bibnamefont {Salvia}}\ and\ \bibinfo {author} {\bibfnamefont {V.}~\bibnamefont {Giovannetti}},\ }\href@noop {} {\bibfield  {journal} {\bibinfo  {journal} {arXiv preprint arXiv:2402.16964}\ } (\bibinfo {year} {2024})}\BibitemShut {NoStop}%
\bibitem [{3-a()}]{3-angles}%
  \BibitemOpen
  \href@noop {} {\bibinfo  {journal} {It should be remarked that the resulting subspace is effectively single-qubit and therefore the most general unitary depends on three parameters (i.e., one more angle). However, we have numerical evidence that considering three parameters yields no substantial differences and, for this reason, we choose to report the simpler ansatz with only two angles to optimize over.}\ }\BibitemShut {NoStop}%
\bibitem [{\citenamefont {{Qiskit contributors}}(2023)}]{Qiskit}%
  \BibitemOpen
\bibfield  {journal} {  }\bibfield  {author} {\bibinfo {author} {\bibnamefont {{Qiskit contributors}}},\ }\href {\doibase 10.5281/zenodo.2573505} {\enquote {\bibinfo {title} {Qiskit: An open-source framework for quantum computing},}\ } (\bibinfo {year} {2023})\BibitemShut {NoStop}%
\bibitem [{\citenamefont {Pedregosa}\ \emph {et~al.}(2011)\citenamefont {Pedregosa}, \citenamefont {Varoquaux}, \citenamefont {Gramfort}, \citenamefont {Michel}, \citenamefont {Thirion}, \citenamefont {Grisel}, \citenamefont {Blondel}, \citenamefont {Prettenhofer}, \citenamefont {Weiss}, \citenamefont {Dubourg}, \citenamefont {Vanderplas}, \citenamefont {Passos}, \citenamefont {Cournapeau}, \citenamefont {Brucher}, \citenamefont {Perrot},\ and\ \citenamefont {Duchesnay}}]{scikit-learn}%
  \BibitemOpen
  \bibfield  {author} {\bibinfo {author} {\bibfnamefont {F.}~\bibnamefont {Pedregosa}}, \bibinfo {author} {\bibfnamefont {G.}~\bibnamefont {Varoquaux}}, \bibinfo {author} {\bibfnamefont {A.}~\bibnamefont {Gramfort}}, \bibinfo {author} {\bibfnamefont {V.}~\bibnamefont {Michel}}, \bibinfo {author} {\bibfnamefont {B.}~\bibnamefont {Thirion}}, \bibinfo {author} {\bibfnamefont {O.}~\bibnamefont {Grisel}}, \bibinfo {author} {\bibfnamefont {M.}~\bibnamefont {Blondel}}, \bibinfo {author} {\bibfnamefont {P.}~\bibnamefont {Prettenhofer}}, \bibinfo {author} {\bibfnamefont {R.}~\bibnamefont {Weiss}}, \bibinfo {author} {\bibfnamefont {V.}~\bibnamefont {Dubourg}}, \bibinfo {author} {\bibfnamefont {J.}~\bibnamefont {Vanderplas}}, \bibinfo {author} {\bibfnamefont {A.}~\bibnamefont {Passos}}, \bibinfo {author} {\bibfnamefont {D.}~\bibnamefont {Cournapeau}}, \bibinfo {author} {\bibfnamefont {M.}~\bibnamefont {Brucher}}, \bibinfo {author} {\bibfnamefont {M.}~\bibnamefont {Perrot}}, \ and\ \bibinfo {author} {\bibfnamefont
  {E.}~\bibnamefont {Duchesnay}},\ }\href@noop {} {\bibfield  {journal} {\bibinfo  {journal} {J. Mach. Learn. Res.}\ }\textbf {\bibinfo {volume} {12}},\ \bibinfo {pages} {2825} (\bibinfo {year} {2011})}\BibitemShut {NoStop}%
\end{thebibliography}%

\section{Acknowledgements}
C.A.P. acknowledges founding from the IQARO (SpIn-orbitronic QuAntum bits in Reconfigurable 2D-Oxides) project of the European Union’s Horizon Europe research and innovation programme under Grant Agreement No. 101115190. C.A.P. acknowledges founding from the PRIN 2022 PNRR project P2022SB73K-Superconductivity in KTaO3 Oxide-2DEG NAnodevices for Topological quantum Applications (SONATA) financed by the European Union–Next Generation EU. G.D.F. and D.F. acknowledge financial support from PNRR MUR Project No. PE0000023-NQSTI. G.D.F. and C.A.P. acknowledge founding from the PRIN 2022 project 2022FLSPAJ ``Taming Noisy Quantum Dynamics" (TANQU). D.J. acknowledges financial support from the Government of Spain (European Union NextGenerationEU PRTR-C17.I1, Severo Ochoa CEX2019-000910-S and TRANQI), Fundació Cellex, Fundació Mir-Puig, and Generalitat de Catalunya (CERCA program).\\
The authors thank A. Ac\'{\i}n for useful discussions at ICFO.
\appendix
\section{Charging two-qubit gate implementation}\label{app:unitary}
We show the decomposition of the charging two-qubit unitary gate $U_{S}^{(c)}$ (as given in Eq.\,\eqref{eq:ucharge}) into single-qubit and two-qubit gates.
\begin{figure}[htbp!]
    \begin{center}
        \includegraphics[scale=0.55]{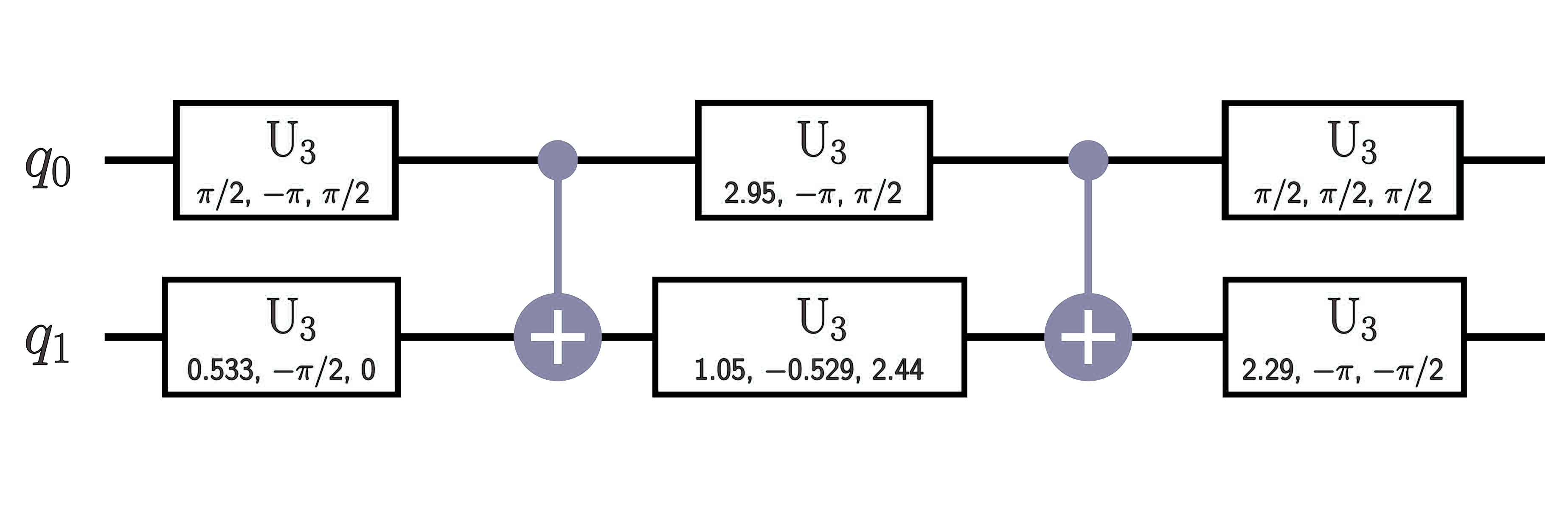}
     
        \caption{\label{fig:qc}Charging gate implementation in terms of single-qubit rotation gates with 3 Euler angles $U_3(\theta, \phi, \lambda)$ and controlled-X two-qubit gates \cite{Qiskit}.} 
    \end{center}
\end{figure}

\section{Quasiprobability distribution approach}
\label{app:variance}
The work extraction at a generic $t\geq 0$ is generated by a time-dependent Hamiltonian driving $F(t)$, which is nonzero only in a time interval of duration $[t,t+\tau]$. At time $t+\tau$, this driving results, in our case, in the application of the local unitary operator to the system's state as described in \eqref{local-ergo}. It should be noticed that in the definition of local ergotropy, $\tau$ is assumed to be arbitrarily small (see e.g. discussions in \citep{castellano2024extended}). Following \cite{francica2022class}, the average work $\langle w \rangle$ can be written as follows,
\begin{equation}
\label{average-work}
    \langle w \rangle = \tr\left(\left[H^{(H)}(\tau)-H(0)\right]\rho_{SE}\right)\,,
\end{equation}
coinciding with $-\mathcal{E_S}(t)$ when the optimal local unitary operator is chosen.
In Eq.\,\eqref{average-work},
$H^{(H)}(t)$ is the Heisenberg representation of the entire Hamiltonian having chosen the time $t$ as initial time. Eq.\,\eqref{average-work} is the first moment of a quasiprobability distribution (see Eq.\,(4) of Ref.\,\cite{francica2022class}) of which we can calculate the moments of any order. In particular, the second moment $\langle w^2 \rangle$ allows us to write the local ergotropy relative fluctuations $\sigma/\mathcal{E}_S$ (or the ones of a lower bound for local ergotropy, when a nonoptimal unitary is selected) at any time, with the work variance given by
\begin{align}
    \sigma^2=\langle w^2 \rangle-\langle w \rangle^2&=\tr\left(\left[H^{(H)}(\tau)-H(0)\right]^2\rho_{SE}\right)+\\
    &-\left\{\tr\left(\left[H^{(H)}(\tau)-H(0)\right]\rho_{SE}\right) \right\}^2\,,
    \nonumber
    \label{variance-t0}
\end{align}
same as Eq.\,\eqref{variance-t} for generic time $t\geq 0$, which at time $t=0$ reduces to $\sigma^2=\bra{\psi^{(c)}}H^2\ket{\psi^{(c)}}-E_c^2$.

\section{Switch-off protocols}
\label{app:so}
For generic $t\geq 0$, we provide explicit expression for the subsystem's ergotropy, $\mathcal{E}_{sub}(t)$ in Eq.\,\eqref{sub-ergo}, for our specific model.
This is achieved by employing the exact formula for a passive two-qubit state \cite{allahverdyan2004maximal}, 
\begin{equation}
\label{eq:ergosyst}
    \mathcal{E}_{sub}(t)= \tr\left({\rho_{S}(t)H_{S}}\right)-\sum_{i=1}^4 r_i^{\downarrow}(t)E_i^{\uparrow}\,.
\end{equation}
The passive state is obtained by computing the eigenvalues $r_i(t)$ of $\rho_{S}(t)$ and ordering them in descending order, multiplied by the ascending order of eigenvalues $E_i$ of $H_{S}$. Furthermore, we account for the switch-off energy price, analyzing $\mathcal{E}_{so}$ in Eq.\,\eqref{so-ergo} incorporating the interaction term $V_{SE}$ with the bath,
\begin{eqnarray}
\label{eq:ergoso}
    &&\mathcal{E}_{so}(t)=\mathcal{E}_{sub}(t)+\\
    &&+\tr\left({\ket{\psi(t)}\bra{\psi(t)}_{SE}(\sigma_z^1+\sigma_z^2)\sum_{i=1}^N \lambda_i (b_i+b^{\dagger}_i)}\right)\,.
    \nonumber
\end{eqnarray}

\section{Variability of optimized ergotropy dataset}\label{app:variability}
\begin{figure}[htbp!]
    \begin{center}
        \includegraphics[scale=0.3]{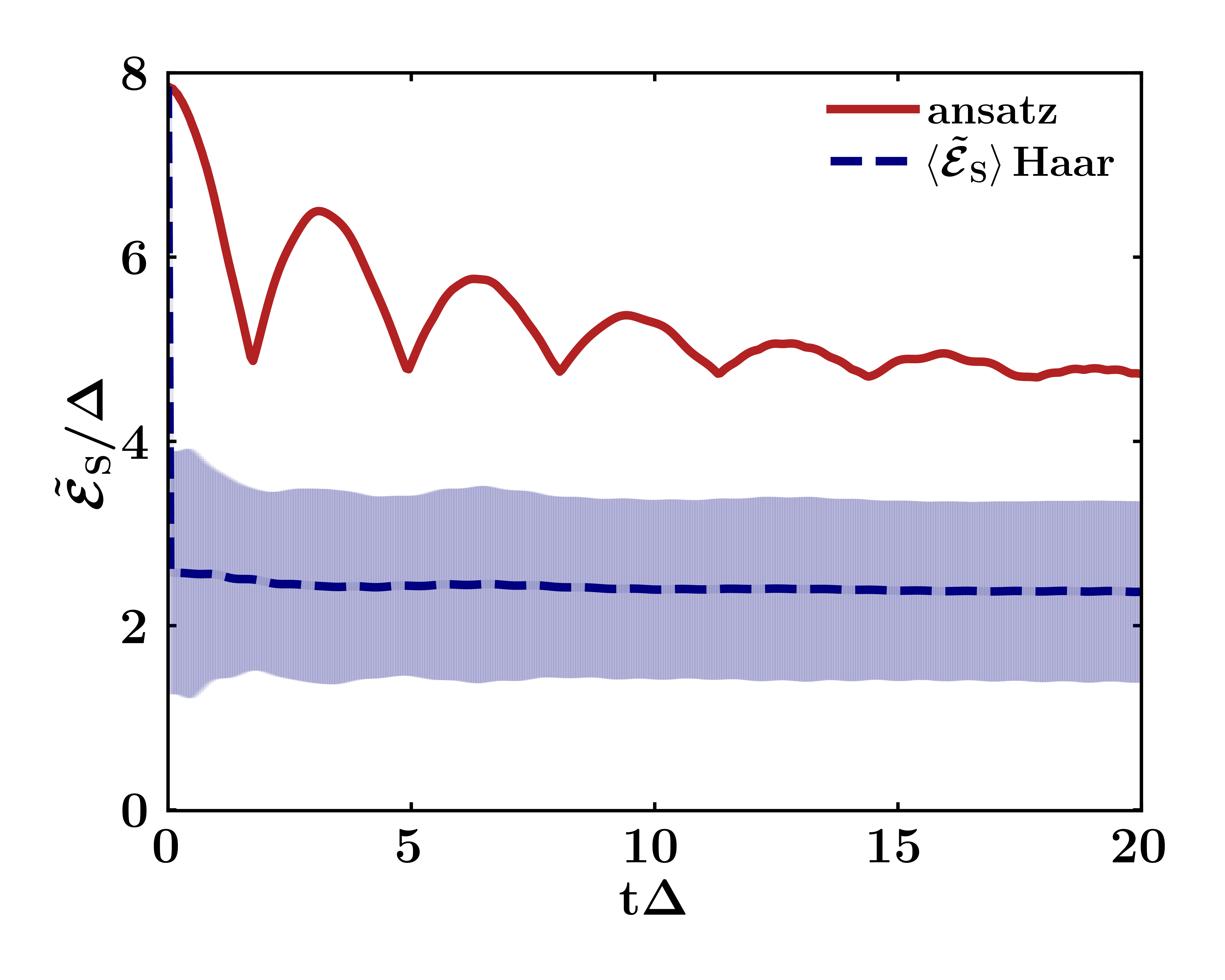}
     
        \caption{\label{fig:ergomean_storage_opt}Lower bound of the local ergotropy in units of $\Delta$ as a function of dimensionless time for $g=0.6/\Delta$ optimized through the ansatz (red solid curve) in \ref{sec:optimization}. Average of the Haar lower bound of the local ergotropy (blue dashed curve). The shaded region represents one standard deviation of the distribution. We employ DMRG to determine the ground state of the $SE$. Then, we apply the unitary operation in \eqref{eq:ucharge} and evolve the charged state using TDVP. We extract work after applying the local unitary gate \eqref{ansatzU} on the MPS for the ansatz and for the optimal parameters $(\bar\theta, \bar\phi)$ found for the Haar distribution.} 
    \end{center}
\end{figure}  
We compare the range of variability in the distribution of Haar matrices used to optimize local work extraction (as discussed in Sec.\,\ref{sec:optimization}) with the lower bound of local ergotropy obtained through the ansatz \eqref{ansatzU} for the unitary operation that extracts work from the system. 

Specifically, in Fig.\,\ref{fig:ergomean_storage_opt},
the average extractable work (another lower bound of local ergotropy) for Haar matrices is depicted as a blue dashed line, while the shaded region represents one standard deviation, indicating the range of variability in the distribution. The red curve corresponding to the ansatz \eqref{ansatzU} lies outside two standard deviations. This observation confirms the quality of the ansatz \eqref{ansatzU} -- already supported by its physical interpretation -- in providing a good lower bound for local ergotropy.  

\section{Bayesian optimization of a two-qubit gate near the ansatz}\label{app:bayesian}
We achieved a slight improvement (approximately $1\%$) in the results discussed in Sec.\,\ref{sec:optimization} by utilizing a unitary operator composed of the one employed in the ansatz (as described in Eq.\,\eqref{ansatzU}) and a generic unitary operator $U=e^{iA}$, where $A$ is Hermitian and can be expressed in terms of Pauli matrices as follows:
\begin{eqnarray}
    A=\sum_{i,j=0}^3 x_{ij} \sigma_i \otimes \sigma_j\,,\\
    \sigma_i \in \{I, \sigma_x, \sigma_y, \sigma_z\}\,.
\end{eqnarray}
This formulation provides us with a parametrization of $U = U({x_{ij}})$. The first term, $x_{00}$, in the decomposition is redundant as it merely introduces a global phase for $U$. We optimize the remaining 15 real variables, $x_{ij} \in [-2, 2]$, using Bayesian optimization with a Gaussian processes regression implemented through Python scikit-learn \cite{scikit-learn}.

\end{document}